\documentclass[a4paper,11pt]{article}

\usepackage{jheppub} 

\usepackage[T1]{fontenc} 
\usepackage[noabbrev]{cleveref}
\usepackage{amsmath,amsfonts,amsbsy,amssymb,array,accents,dsfont}
\usepackage{enumerate,latexsym,graphicx}
\usepackage{xcolor,tikz}
\usepackage{tcolorbox}
\usepackage{mathtools}

\usepackage{subfigure}
\usepackage{stmaryrd}
\usepackage[compat=1.1.0]{tikz-feynman}

  \def\cC{\mathcal {C}}

\def\cV{\mathcal {V}} \def\cW{\mathcal {W}}

\newcommand{\beq}{\begin{equation}}
\newcommand{\eeq}{\end{equation}}
\newcommand{\bea}{\begin{eqnarray}}
\newcommand{\eea}{\end{eqnarray}}

\newcommand{\vep}{\varepsilon}
\newcommand{\la}{\lambda}

\newcommand{\der}{\partial}

\newcommand{\nn}{\nonumber}

\renewcommand{\>}{\rangle}
\renewcommand{\Re}{{\rm Re}}
\newcommand{\N}{{\cal{N}}}

\usepackage{feynmp-auto}

\usepackage{braket}
\usepackage{tikz}
\usetikzlibrary{matrix,calc,positioning,decorations.markings,decorations.pathmorphing,decorations.pathreplacing}
\usetikzlibrary{arrows,cd}
\usepackage{bbding}
\usetikzlibrary{positioning}
\tikzset{>=stealth}

\newcommand{\psisl}{\psi}
\newcommand{\psisu}{\chi}

\newcommand{\del}{\partial}

\newcommand{\qqquad}{\;, \quad\qquad}  

\newcommand{\CC}{\mathbb{C}}
\newcommand{\RR}{\mathbb{R}}

\newcommand{\NN}{\mathbb{N}}

\newcommand{\VV}{\mathbb{V}}

\newcommand{\Aa}{\mathcal{A}}
\newcommand{\Bb}{\mathcal{B}}
\newcommand{\Cc}{\mathcal{C}}
\newcommand{\Dd}{\mathcal{D}}

\newcommand{\Ff}{\mathcal{F}}

\newcommand{\Ii}{\mathcal{I}}
\newcommand{\Jj}{\mathcal{J}}

\newcommand{\Nn}{\mathcal{N}}
\newcommand{\Oo}{\mathcal{O}}

\newcommand{\Vv}{\mathcal{V}}

\newcommand{\Ww}{\mathcal{W}}

\newcommand{\R}{\ensuremath{\mathbb{R}}}

\newcommand{\zb}{\bar{z}}
\newcommand{\xb}{\bar{x}}

\newcommand{\w}{\omega}

\newcommand{\ab}{{\bar{a}}}
\newcommand{\bb}{{\bar{b}}}
\newcommand{\cb}{{\bar{c}}}
\newcommand{\ub}{{\bar{u}}}

\usepackage{color}

\usepackage[normalem]{ulem}  

\definecolor{darkred}{rgb}{0.6,0,0}
\definecolor{darkblue}{rgb}{0,0,0.6}

\newcommand{\be}{\begin{equation}}
\newcommand{\ee}{\end{equation}}

\usepackage[bbgreekl]{mathbbol}
\usepackage{amsfonts}
\DeclareSymbolFontAlphabet{\mathbb}{AMSb} 
\DeclareSymbolFontAlphabet{\mathbbl}{bbold} 

\newcommand{\ww}{\boldsymbol{\w}}
\newcommand{\jj}{\mathrm{j}}

\ExplSyntaxOn

\NewDocumentCommand{\pFq}{O{}mmmmm}
 {
  \group_begin:
  \keys_set:nn { hypergeometric } { #1 }
  \hypergeometric_print:nnnnn { #2 } { #3 } { #4 } { #5 } { #6 }
  \group_end:
 }
\NewDocumentCommand{\hypergeometricsetup}{m}
 {
  \keys_set:nn { hypergeometric } { #1 }
 }

\tl_new:N \l_hypergeometric_divider_tl
\tl_new:N \l_hypergeometric_left_tl
\tl_new:N \l_hypergeometric_right_tl

\keys_define:nn { hypergeometric }
 {
  symbol .tl_set:N = \l_hypergeometric_symbol_tl,
  symbol .initial:n = F,
  separator .tl_set:N = \l_hypergeometric_separator_tl,
  separator .initial:n = {},
  skip .tl_set:N = \l_hypergeometric_skip_tl,
  skip .initial:n = 8,
  divider .choice:,
  divider/semicolon .code:n = \tl_set:Nn \l_hypergeometric_divider_tl { \;; },
  divider/bar .code:n = \tl_set:Nn \l_hypergeometric_divider_tl { \;\middle|\; },
  divider .initial:n = semicolon,
  fences .choice:,
  fences/brack .code:n = 
   \tl_set:Nn \l_hypergeometric_left_tl {[}
   \tl_set:Nn \l_hypergeometric_right_tl {]},
  fences/parens .code:n = 
   \tl_set:Nn \l_hypergeometric_left_tl {(}
   \tl_set:Nn \l_hypergeometric_right_tl {)},
  fences .initial:n = brack,
 }

\cs_new_protected:Nn \hypergeometric_print:nnnnn
 {
  {} \sb {#1} \l_hypergeometric_symbol_tl \sb { #2 }
  \left\l_hypergeometric_left_tl
  \genfrac .. 
           {0pt} 
           {} 
           { \__hypergeometric_process:n { #3 } } 
           { \__hypergeometric_process:n { #4 } } 
  \l_hypergeometric_divider_tl
  #5
  \right\l_hypergeometric_right_tl
 }

\cs_new_protected:Nn \__hypergeometric_process:n
 {
  \clist_use:nn { #1 }
   {
    {\l_hypergeometric_separator_tl}
    \mspace { \l_hypergeometric_skip_tl mu }
   }
 }

\ExplSyntaxOff

\title{\boldmath Superstring four-point functions in  AdS$_3\times S^3\times T^4$  
}

\author[a]{Emiliano Barone}
\author[a]{, Sergio Iguri}
\author[b]{, Nicolas Kovensky}
\author[a,c]{and Juli\'an H.~Toro}

\affiliation[a]{Instituto de Astronomía y Física del Espacio (IAFE), CONICET - Universidad de Buenos Aires, Facultad de Ciencias Exactas y Naturales,  Ciudad Universitaria, 1428 Buenos Aires, Argentina.}
\affiliation[b]{Instituto de F\'isica de La Plata - CONICET
Diagonal 113 e/ 63 y 64, 1900 - La Plata, Argentina.}
\affiliation[c]{Instituto de Investigaciones Matemáticas Luis A. Santaló (IMAS), CONICET - Universidad de Buenos Aires, Facultad de Ciencias Exactas y Naturales, Ciudad Universitaria, 1428 Buenos Aires, Argentina.}
\emailAdd{siguri@iafe.uba.ar}
\emailAdd{nicolas.kovensky@ipht.fr}
\emailAdd{jtoro@dm.uba.ar}
\emailAdd{ebarone@iafe.uba.ar}

\abstract{ We study spacetime four-point functions of chiral primary operators  for superstrings propagating in AdS$_3\times S^3\times T^4$ from the worldsheet perspective, allowing for external states with arbitrary spectral flow charges. We work at small boundary cross-ratio and consider both extremal (supersymmetry-protected) and non-extremal configurations. 
The string correlators involve a non-trivial integration over the worldsheet moduli, and we demonstrate how the exchange of short and long strings is encoded in distinct integration regions. We interpret our results in terms of the operator product expansion in the boundary theory, and make explicit the precise matching with the available data for the relevant holographic correlators. 
}

\newcommand{\of}[1]{\left(#1\right)}
\newcommand{\off}[1]{\left[#1\right]}

\begin{document} 
\hypergeometricsetup{
  separator={,},
  divider=bar,
}
\maketitle
\flushbottom

\section{Introduction}
\label{sec: intro}

As is well-known, four-point functions in conformal field theories (CFTs) are severely constrained: they can only depend on the cross-ratios, and must be consistent with the operator product expansion (OPE) and crossing symmetry. Nevertheless, these observables are highly non-trivial in most interacting models, and, except for some exceptional examples, they are not known in closed form. This remains true in the two-dimensional case, where the symmetry algebra becomes infinite-dimensional, and even under the additional assumption of supersymmetry. 

The AdS/CFT correspondence \cite{Maldacena:1997re} establishes a relation between gravity/strings in asymptotically anti-de Sitter (AdS) spacetimes -- the bulk -- and conformal field theories defined on their boundary. On a more fundamental level, this is a duality between the two-dimensional  worldsheet theory describing string propagation in an asymptotically AdS$_{d+1}$ background, and a $d$-dimensional \textit{holographic} CFT (HCFT). Hence, a natural question that arises is how the OPE structure and four-point functions of the holographic CFT  could be derived from those of the worldsheet model.

Dual pairs where both the worldsheet and boundary CFTs are non-topological models under exact control are extremely rare, but they do exist. The most studied and best understood example corresponds to the duality between strings in an AdS$_3\times S^3 \times T^4$ supported by Neveu-Schwarz (NS) flux, and a particular deformation of a two-dimensional  superconformal  field theory (SCFT) of the symmetric orbifold type. Surprisingly, the question raised above is,  so far, not well understood even in this simplified context, where, for instance, some correlators of chiral primary operators are not renormalized \cite{Baggio:2012rr,deBoer:2008ss}. In other words, the precise relation between the worldsheet and spacetime OPEs is yet to be fully established. 

In this paper we address such questions by describing how the structure of holographic four-point functions at large $N$ is derived from string theory as defined by the $\N=1$ Wess-Zumino-Witten (WZW) model on SL(2,$\R$)$\times$SU(2)$\times$U(1)$^4$. Although this worldsheet theory is believed to be solvable, the explicit  computation of correlation functions is technically challenging, in particular when spectrally flowed operators are included \cite{Maldacena:2000hw,Maldacena:2001km}. These vertex operators have a complicated OPEs with the symmetry currents of the model, which renders the derivation of the corresponding correlators quite involved already at the bosonic level \cite{Giribet:2000fy,Maldacena:2001km,Ribault:2005ms,Minces:2007td,Iguri:2009cf,Cagnacci:2015pka}. Additional issues arise in the supersymmetric case mainly due to the interplay between spectral flow, the presence of fermionic fields, and the picture-changing procedure \cite{Pakman:2007hn,Giribet:2007wp,Iguri:2022pbp,Iguri:2023khc}. Moreover, in order to derive holographic data from these observables, one must integrate over the worldsheet moduli, which is automatic for two- and three-point functions, but considerably more intricate for four-point functions \cite{Maldacena:2001km,Cardona:2010qf}.

Over the last few years, the insights of \cite{Eberhardt:2019ywk} gave rise to a set of interesting new techniques for dealing with spectral flow. This led, for example, to the derivation of the holographic correspondence in the limit where strings become tensionless, i.e.~when the geometry is sourced by a single NS5 brane together with  $n_1$ fundamental strings \cite{Gaberdiel:2017oqg,Giribet:2018ada}. At this point in moduli space the AdS$_3$ becomes string-size, and the relevant super WZW model on (the universal cover of) SL(2,$\R$) has level $n_5=1$. In this somewhat simplified limit, where the worldsheet spectrum is built only from long strings, the dual was identified as the \textit{undeformed} symmetric orbifold involving $n_1$ copies of the $T^4$ CFT \cite{Eberhardt:2018ouy,Eberhardt:2020bgq,Dei:2020zui,Dei:2023ivl,Eberhardt:2025sbi}.

For $n_5 > 1$ the story is considerably more involved. Nevertheless, the local Ward identities of \cite{Eberhardt:2019ywk} were used in \cite{Dei:2021xgh,Bufalini:2022toj} to derive all bosonic two- and three-point point functions involving vertex operators with arbitrary spectral flow charges exactly, albeit in integral form. This was later extended to the supersymmetric case, which allowed for a full derivation\footnote{To be precise, a small family of operators are not included in this analysis, see \cite{Seiberg:1999xz}.} of the spacetime chiral ring from the worldsheet perspective \cite{Iguri:2022pbp,Iguri:2023khc}. A concrete, perturbative proposal for the holographic CFT dual to bosonic strings in AdS$_3$ was put forward in \cite{Eberhardt:2021vsx,Dei:2022pkr}, see also \cite{Balthazar:2021xeh}. This was further tested recently in \cite{Knighton:2023mhq,Knighton:2024qxd} by means of free field methods, which accurately describe the near-boundary regime of the bulk theory\footnote{See \cite{Sriprachyakul:2025ubx} for a recent application to  the spacetime OPE from the string perspective.}. For the supersymmetric case, the situation is slightly less conclusive  \cite{Sriprachyakul:2024gyl,Yu:2024kxr,Yu:2025qnw} 

Here we build on these results in order to study spacetime four-point functions for strings in AdS$_3\times S^3\times T^4$. Our analysis is based on the bosonic formula proposed in \cite{Dei:2021yom}. Significant evidence for its validity  was obtained recently in \cite{Iguri:2024yhb} by studying the factorization properties of such correlators. We extend this analysis to the supersymmetric context, and further discuss in detail the role of chiral primary states. More explicitly, we consider four-point functions with insertions given by the worldsheet avatars of the chiral operators in arbitrary twisted sectors of the boundary theory at genus zero, both extremal and non-extremal. Following \cite{Maldacena:2001km}, we carry out the integration over the worldsheet moduli and derive all relevant single-cycle exchange channels involving both short and long strings. In all cases, we show that the factorization properties of supersymmetric worldsheet four-point correlators are consistent with the structure constants and selection rules of the worldsheet theory \cite{Dei:2019osr,Bufalini:2022toj}, and that they reproduce exactly all known results for the spacetime OPE of chiral primary operators \cite{Pakman:2009zz,Pakman:2009ab}. This holds not only for protected quantities, but also for a family of non-extremal correlators, of which a small, unflowed subset was briefly discussed in \cite{Cardona:2010qf}.

Our results provide additional evidence for the validity of the conjecture put forward in \cite{Dei:2021yom}, and suggest a formal proof should be within reach. This formula should be understood as an integral relation between flowed four-point functions and their unflowed cousins. Even though the latter are not known in closed form, we argued in \cite{Iguri:2024yhb}, and confirm in the present work, that this integral transform is of practical use. In short, when combined with known results about the behaviour of the unflowed SL(2,$\R$) conformal blocks in different regions of parameter space, the integral expression for flowed correlators enables the study of both spectrally flowed conformal blocks and spacetime four-point functions. Moreover, we show that, similarly to what was done in \cite{Iguri:2023khc}, it can also be used to describe the non-trivial contributions from the fermionic fields and the SU(2) sector of the worldsheet theory. Finally, we expect our results to shed light on the precise definition of the holographic CFT. 

This paper is organized as follows. Section \ref{sec: setup} describes the basic setup and provides a review of the techniques and known results that will be used throughout the paper.  
In Section \ref{Sec: On the bosonic four-point function} we consider a simplified situation, and discuss four-point functions of lowest-weight operators for bosonic strings in AdS$_3$. We describe some subtleties related to the integration over the worldsheet moduli, and analyze these integrated correlators in terms of the factorization properties of the spacetime theory, i.e.~the OPE expansion of the holographic CFT. We show that, using the formula of \cite{Dei:2021yom} and further extending the analysis of \cite{Iguri:2024yhb}, it is possible to describe consistently all allowed exchanges involving both short and long strings with different spectral flow charges.

Section \ref{sec: Extremal four-point functions of chiral primary operators}
contains the main results of this paper. We show how the methods employed in the bosonic case can be extended to the supersymmetric model. First we consider the short string contributions to the extremal four-point functions of chiral primary states. We show that, for arbitrary spectral flow charges satisfying the extremality condition, the worldsheet results matches the holographic computation of \cite{Pakman:2009ab} obtained from the symmetric orbifold theory exactly (at large $N$). Along the way, we also rederive the results of \cite{Iguri:2023khc} from a complementary perspective. Given that these correlators are protected by supersymmetry and hence not renormalized, this constitutes an important consistency check for our  procedure, and a non-trivial extension of the results of \cite{Cardona:2010qf}, where only the unflowed sector was considered. We then consider processes with intermediate channels associated with long strings, and show that, after integrating over the worldsheet cross-ratio, all results derived from the string description are consistent with the conformal block expansion of the boundary theory.

Finally, in Section \ref{sec: non-extremal} we discuss a specific family  of non-extremal, and hence, in principle, non-protected correlators. We show that, as for the unflowed cases considered in \cite{Cardona:2010qf}, for all relevant flowed sectors one obtains a precise matching with the symmetric orbifold predictions also in these cases. This suggests that such correlators should also be protected by supersymmetry somehow, although this has not been completely established so far. 
We end the paper by summarizing our results and presenting the outlook in section \ref{sec: conclusions}. A small number of useful formulas and auxiliary computations are relegated to the appendices.

\section{Setup, conventions and previous results}
\label{sec: setup}

The dynamics of string propagation in an AdS$_3 \times$S$^3 \times$T$^4$ background with NSNS fluxes is described by the $\N=1$ supersymmetric WZW model on SL(2,$\mathbb R$)$\times$SU(2)$\times$U(1)$^4$. In this section, we review the different sectors of the model, largely following the notational conventions introduced in \cite{Iguri:2022pbp}. 

\subsection{Basic definitions}

The affine SL(2,$\R$) and SU(2) holomorphic currents and fermions are denoted as $J^{a}$, $\psi^a$ and $K^a$, $\chi^a$, respectively, with $a=0,1,2$. They satisfy the following OPEs: 
\begin{alignat}{4}
&J^a(z)J^b(w) &&\sim \frac{n_5\eta^{ab}}{2(z-w)^2}+\frac{i\vep^{ab}{}_{c}J^c(w)}{z-w},\quad&&K^a(z)K^b(w) &&\sim \frac{n_5\delta^{ab}}{2(z-w)^2}+\frac{i\vep^{ab}{}_{c}K^c(w)}{z-w},\\
&J^a(z)\psi^b(w)&&\sim \frac{i\vep^{ab}{}_{c}\psi^c(w)}{z-w},\quad &&K^a(z)\chi^b(w)&&\sim \frac{i\vep^{ab}{}_{c}\chi^c(w)}{z-w},\\
&\psi^a(z)\psi^b(w)&&\sim \frac{n_5\eta^{ab}}{2(z-w)},\quad &&\chi^a(z)\chi^b(w)&&\sim \frac{n_5\delta^{ab}}{2(z-w)},
\end{alignat}
where $n_5$, the number of NS5-branes sourcing the geometry, is the level in both cases, while $\vep^{012}=1$, $\eta^{ab} = \eta_{ab} = (-++)$ and $\delta^{ab}=\delta_{ab} = (+++)$. Similar expressions hold for the antiholomorphic currents. By means of the holographic correspondence, the zero modes of the SL(2,$\R$) currents $J^a_0$ are identified with the global modes of the spacetime Virasoro algebra, while those of the SU(2) currents $K^a_0$ are identified with the R-symmetry generators of the spacetime $\N=(4,4)$  supersymmetry algebra.  

Ladder operators are defined as $J^\pm = J^1 \pm i J^2$, and $K^\pm$, $\psi^\pm$ and $\chi^\pm$ are defined analogously. 
The supersymmetric currents can be decomposed as 
\beq
J^a = j^a + \hat{\jmath}^a\qqquad
K^a = k^a + \hat{k}^a.
\eeq
Here the $j^a$ generate a bosonic SL(2,$\R$)$_{k}$ subalgebra with a shifted level $k = n_5 + 2$. The same goes for the SU(2)$_{k'}$ associated to the bosonic currents $k^a$, for which the level is $k' = n_5-2$, hence the RNS formalism employed here is most useful for $n_5 \geq 2$, which excludes the tensionless case. On the other hand, the fermionic currents  
\beq 
\hat{\jmath}^a = -\frac{i}{n_5}\vep^{a}{}_{bc}\psi^b \psi^c\qqquad 
\hat{k}^a = -\frac{i}{n_5}\vep^{a}{}_{bc}\chi^b \chi^c \,, \label{Fermionic Currents}
\eeq
can be understood in terms of SL(2,$\R$)$_{-2}$ and SU(2)$_{2}$ algebras, which decouple from the bosonic sector. 
The (matter) stress tensor $T$ and supercurrent $G$ can be  expressed  as 
\begin{eqnarray}
    T &=& \frac{1}{n_5} \left(j^a j_a - \psisl^a \der \psisl_a + 
    k^a k_a - \psisu^a \der \psisu_a 
    \right) + T_{\rm torus
    } \,,  
    \label{TAdS3S3T4def}
    \\
    G &=& \frac{2}{n_5} \left(
    \psisl^a j_a + \frac{2i}{n_5}\psisl^0 \psisl^1 \psisl^2 + 
    \psisu^a k_a - \frac{2i}{n_5}\psisu^0 \psisu^1 \psisu^2
    \right) + G_{\rm torus} \, .
    \label{GAdS3S3T4def}
\end{eqnarray}
After including the $bc$ and $\beta \gamma$ ghost systems, one obtains the BRST charge 
\begin{equation}
\label{eq:BRSToperator}
    {\cal{Q}} = \oint dz \left[ c \left(T + T_{\beta\gamma}\right) - \gamma \, G + c(\der c) b - \frac{1}{4} b \gamma^2\right] \, .
\end{equation}
As usual, we also define $\beta = e^{-\varphi} \del \xi$ and $\gamma = \eta \:\! e^{\varphi}$, with 
 $\xi(z)\eta(w) \sim (z-w)^{-1}$, such that $\varphi$ has a background charge $Q_\varphi = -2$.

\subsection{The bosonic SL$(2,\RR)$ and SU$(2)$ models} 

The physical states of the $\N=1$ worldsheet SCFT are classified according to the bosonic SL$(2,\mathbb{R})$ representations they belong to \cite{Giveon:1998ns,Kutasov:1999xu,Maldacena:2000hw,Maldacena:2001km}. One must distinguish between states in the spectrally flowed continuous representations $C_{j}^{\alpha,\w}$ and those in the (possibly flowed) discrete lowest/highest-weight representations  $D_{j}^{\pm,\w}$. These correspond to long and short strings, respectively. The relevant quantum numbers are the spectral flow charge $\w \in \mathbb{Z}$, the unflowed SL$(2,\RR)$ spin $j$, and for the continuous case there is an additional parameter $\alpha \in [0,1)$ related to the eigenvalues of $J_0^3$. For short strings, the unflowed spin is real, the allowed range $\frac{1}{2} < j < \frac{k-1}{2}$ being determined by normalizability and no-ghost theorems, while for long strings $j$ takes values in $\frac{1}{2} + i \R$. 
In particular, supergravity states belong to ${\cal{D}}_j^{\pm} \equiv {\cal{D}}_j^{\pm,0}$. Although such   states can be understood via analytic continuation from the $H_3^+$ model explored in \cite{Teschner:1997ft,Teschner:1999ug}, the inclusion of spectrally flowed states is essential for maintaining the consistency of the worldsheet spectrum. 

Vertex operators defined in the so-called $x$-basis are denoted $V_{jh}^{\w}(x,z)$, where $x$ and $z$ are the boundary and worldsheet complex coordinates, and one can restrict to $\w\geq 0$ \cite{Maldacena:2001km}. The spacetime conformal weight $h=m + \frac{k}{2}\w$ is defined by the $j^3_0$ eigenvalue $m$ before spectral flow. In the unflowed sector $h=j$, hence operators with $\w=0$ are written as $V_j(x,z)$. Even though all these operators also depend on the antiholomorphic variables $\bar{z}, \bar{x}$ (and should have an additional label $\bar{h}$) we will omit them throughout the paper.
Actually, not all flowed representations are independent. In the discrete sector, one has the series identifications
\begin{equation}
\label{series-identif}
    V_{j,h=j + \frac{k}{2}\w}^\w(x,z) = {\cal{N}}(j) V_{\frac{k}{2}-j,h}^{\w+1}(x,z) \, , \qquad 
    {\cal{N}}(j) = \sqrt{\frac{B(j)}{B\left(
    \frac{k}{2}-j\right)}} \, ,
\end{equation}
where
\begin{equation}
    B(j)=\frac{2j-1}{\pi}
    \frac{\Gamma[1-b^2(2j-1)]}{\Gamma[1+b^2(2j-1)]} \, \nu^{1-2j} \, , \quad  b^2 = (k-2)^{-1}\,,
     \label{B(j)} \,
\end{equation} 
with $\nu$ a free parameter of the theory, which we ignore for now.
This relates lowest-weight states of spin $j$ and spectral flow charge $\w$ and highest-weight states of spin $\frac{k}{2}-j$ and spectral flow charge $\w+1$, and hence translates into the equivalence $\Dd^{+,\w}_j\equiv  \Dd^{-,\w+1}_{\frac{k}{2}-j}$. The continuous sector further exhibits a reflection symmetry $j \leftrightarrow 1-j$. In particular, when $\w=0$ one has 
\begin{equation}
\label{reflection w=0}
V_{j}(x,z) = B(j)\int d^2x \, |x-x'|^{-4j} V_{1-j}(x',z) \, ,
\end{equation}
where the integral is performed over the full complex plane. For $\w>0$, on the other hand,  we have
\begin{equation}
\label{reflection w>0}
    V_{jh}^\w(x,z) = R(j,h,\w) V_{1-j,h}^\w(x,z) \, ,
\end{equation}
with 
\begin{equation}
\label{def N and R}
         R(j,h,\w) = \frac{ \pi \gamma \left(h-\frac{k}{2}\w+j\right) B(j)}{\gamma(2j) \gamma\left(h-\frac{k}{2}\w+1-j\right)} \, , \qquad 
        \gamma(x) = \frac{\Gamma(x)}{\Gamma(1-\bar{x})} \, .
\end{equation}

The worldsheet Virasoro algebra is derived using the Sugawara construction, and, for instance, the bosonic SL(2,$\R$)$_k$ sector yields a central charge $c=\frac{3k}{k-2}$. Unlike the unflowed states, vertex operators $V_{jh}^{\w}(x,z)$ with $\w>0$ are not affine primaries, but they are (worldsheet) Virasoro primaries of weight
\begin{equation}
\label{def Delta w}
    \Delta^\w = -\frac{j(j-1)}{k-2} - h\w + \frac{k}{4}\w^2\,.
\end{equation}
Is has proven useful to combine the above vertex operators in terms the so-called $y$-basis \cite{Dei:2019osr}, see also \cite{Iguri:2022eat}. The relevant definition is given by the Mellin-type transform\footnote{Though we write absolute values squared throughout this paper, this is, strictly speaking, an abuse of notation: $\bar{h}$ is not necessarily the complex conjugate of $h$.}
\begin{equation}
    V_{jh}^{\w}(x,z) = 
    \int d^2y \,  \big|y^{j+\frac{k}{2}\w-h-1}\big|^2  
    V_j^\w (x,y,z)\,.
    \label{xtoybasis}
\end{equation}
This leads to the following OPEs with the currents of the theory:
\begin{eqnarray}
   j^+(w)  V^{\w}_{j}(x,y,z)  &=&  \frac{\der_y 
    V_{j}^\w (x,y,z) }{(w-z)^{\w+1}} + 
   \sum_{n=2}^{\w} \frac{\left(J_{n-1}^+ 
    V_{j}^\w\right) (x,y,z) }{(w-z)^n}
    +\frac{\der_x 
    V_{j}^\w (x,y,z)  
      }{(w-z)}  + \cdots \, , \nn \\
      j^3(x,w)  V^{\w}_{j}(x,y,z)  &=& 
   \frac{\left(y \der_y + j + \frac{k}{2}\w\right)
    V_{j}^\w (x,y,z)  
      }{(w-z)}  + \cdots \, , \\
      j^-(x,w)  V^{\w}_{j}(x,y,z)  &=&  (w-z)^{\w-1} \left(y^2\der_y + 2 j y\right)
    V_{j}^\w (x,y,z) + \cdots \, , \nn
\end{eqnarray}
where we have introduced 
\begin{equation}
    j^3(x,z) = j^3(z) - x j^+(z) \, , \qquad 
    j^-(x,z) = j^-(z) - 2 x j^3(z) + x^2 j^+(z) \, . 
\end{equation}
For $y$-basis operators the reflection symmetry reads 
\begin{equation}
\label{reflection ybasis}
V^\w_{j}(x,y,z) = B(j)\int d^2y' \, |y-y'|^{-4j} V^\w_{1-j}(x,y',z) \, .
\end{equation}
Local Ward identities \cite{Eberhardt:2019ywk} impose strong constraints on correlation functions. The latter arise from the vanishing of the first $\w-1$ terms in the expansion of the product $j^-(x,w)  V^{\w}_{j}(x,y,z)$ around $z=w$ \cite{Eberhardt:2019ywk}, and can be recast in terms of differential equations for the dependence on $y$. 

While so far we have discussed the bosonic SL(2,$\mathbb R$) sector, analogous results apply to the bosonic SU(2) sector of the theory. Using the isospin variable $u$, the corresponding bosonic primaries $W_l(u,z)$ possess integer and half-integer spins bounded $0\leq l \leq k'/2$, and the associated modes $W_{l,n}(z)$ have spin projections $n=-l,-l+1, \dots,l-1,l$ \cite{Zamolodchikov:1986bd}. The spectrally flowed counterparts, denoted by $W^{\w}_{l,n}(u,z)$, are lowest-weight states, and their flowed spin is given by $l_\w \equiv -n+\frac{k'}{2} \w$.

\subsection{Spectrally-flowed correlators in the bosonic SL(2,$\R$) WZW model}

Exact correlation functions in the bosonic SL$(2,\mathbb{R})$ model were first computed for unflowed cases in \cite{Maldacena:2001km} via analytic continuation from the H$_3^+$ theory \cite{Teschner:1999ug}. Recent progress for the flowed cases was made in  \cite{Eberhardt:2019ywk,Dei:2021xgh,Dei:2021yom,Iguri:2022eat,Bufalini:2022toj,Iguri:2023khc}.

\subsubsection{Two- and three-point functions}

The two-point functions read \cite{Maldacena:2001km} 
\begin{align}
    & 
    \langle V^{\w_1}_{j_1 h_1}(0,0)V^{\w_2}_{j_2 h_2}(\infty,\infty)\rangle =  \delta_{\w_1,\w_2} \delta^{(2)}(h_1-h_2)
    \off{\delta(j_1-j_2)R(j_1,h_1,\w_1)
    +\delta(j_1+j_2-1)} \, .
\end{align}
Using the integral transform \eqref{xtoybasis}, this can be derived from the corresponding result for $y$-basis operators \cite{Dei:2021xgh}, namely  \begin{align}
    \langle V^{\w_1}_{j_1}(0,y_1,0)V^{\w_2}_{j_2}(\infty,y_2,\infty)\rangle = & \, \delta_{\w_1,\w_2}B(j_1)\delta(j_1-j_2) |1-y_1y_2|^{-4j_1}\\
    &+\delta_{\w_1,\w_2}\delta(j_1+j_2-1)|y_1|^{-2j_1}|y_2|^{-2j_2}\delta^{(2)}(1-y_1y_2) \, \, . \nn
\end{align}
For $n$-point correlation functions with $n>2$, the spectral flow charge is not a conserved quantity, although one has the constraint \cite{Maldacena:2001km} 
\begin{equation}  
2\max(\w_i)\leq \sum_{i=1}^n \w_i + 1 \, .\label{Wselectionrules}
\end{equation} 
The exact expression for generic $y$-basis three-point functions was conjectured in \cite{Dei:2021xgh} and later established in \cite{Bufalini:2022toj}. It takes the form 
\begin{align}
       &\langle V_{j_1}^{\w_1}(y_1)V_{j_2}^{\w_2}(y_2)V_{j_3}^{\w_3}(y_3)\rangle \equiv \left\langle V^{\w_1}_{j_1}(0, y_1, 0) \,  V^{\w_2}_{j_2}(1, y_2, 1) \, V^{\w_3}_{j_3}(\infty, y_3, \infty) \right\rangle=\nn\\
       &\hspace{4cm}\left\{
\begin{array}{cc}
     C_{-}(j_1,j_2,j_3) \Big| Z_{123}^{\frac{k}{2}-J} \prod_{i=1}^3 Z_i^{J-\frac{k}{2}-2j_i}\Big|^2 \,, & 
     \qquad {\rm for} \,\, \w \,\,  {\rm odd}\\[1ex]
     C_{+}(j_1,j_2,j_3) \Big| Z_{\emptyset}^{J-k} \prod_{i<\ell}^3 Z_{i\ell}^{J-2j_i-2j_\ell}\Big|^2 \,, & 
     \qquad {\rm for} \,\, \w \,\,  {\rm even}
\end{array}
      \right.
      \label{3pt-final}
\end{align}
where $J=j_1+j_2+j_3$ and $\w = \w_1+\w_2+\w_3$. 
For any subset $I \subset \{ 1,2,3 \}$, the functions $Z_I$, which play the role of \textit{generalized differences}, are defined as\footnote{For later convenience, we keep the original notation $(Z_I,Q_{\boldsymbol{\w}}) \to (X_I,P_{\boldsymbol{\w}})$ for four-point functions. }
\begin{equation}
   Z_I(y_1,y_2,y_3)\equiv \sum_{i \in I:\ \varepsilon_i=\pm 1} Q_{\boldsymbol{\w}+\sum_{i \in I} \varepsilon_i e_{i}} \prod_{i\in I} y_i^{\frac{1-\varepsilon_i}{2}} \ . 
\label{Z_I-3pt} 
\end{equation} 
with $\boldsymbol{\w}=(\w_1, \w_2, \w_3)$, $
e_1 = (1,0,0)$, $e_2 = (0,1,0)$, and $e_3 = (0,0,1)$. The numbers $Q_{\boldsymbol{\w}}$ are given by
\begin{equation}
    Q_{\boldsymbol{\w}} = 0 \qquad \text{if} \qquad \sum_j \w_j < 2 \max_{i=1,2,3} \w_i \quad \text{or}\quad \sum_i \w_i \in 2\mathds{Z}+1 \, , 
\end{equation} 
while otherwise 
\begin{equation}
  Q_{\boldsymbol{\w}} =S_{\boldsymbol{\w}} \frac{G\left(\frac{-\w_1+\w_2+\w_3}{2} +1\right) G\left(\frac{\w_1-\w_2+\w_3}{2} +1\right) G\left(\frac{\w_1+\w_2-\w_3}{2} +1\right) G\left(\frac{\w_1+\w_2+\w_3}{2}+1\right)}{G(\w_1+1) G(\w_2+1) G(\w_3+1)}  \ .  
\label{Qw-definition}  
\end{equation}
Here, $G(n)=\prod_i^{n-1}\Gamma(i)$ denotes the Barnes $G$-function, and $S_\w$ is an unimportant phase. The overall factor $C_{+}(j_1,j_2,j_3)$ equals the unflowed structure constant $C(j_1,j_2,j_3)$ given in \cite{Teschner:1997ft,Teschner:1999ug}, while $C_{-}(j_1,j_2,j_3)$ is defined as
\begin{equation}
\label{consdei}
    C_{-}(j_1,j_2,j_3) = 
    {\cal{N}}(j_3) C\left(j_1,j_2,\frac{k}{2}-j_3 \right) \, .
\end{equation}  
The reflection symmetry inherited from \eqref{reflection w=0} implies the identity
\begin{equation}
    \frac{\pi\gamma(j-j_3+j_2)\gamma(j+j_3-j_2)C(1-j,j_2,j_3)}{\gamma(2j)  } \label{Creflection}=
     \frac{C(j,j_2,j_3)}{B(j)}\,.
\end{equation}
We provide the explicit form of the three-point functions for the specific cases where $\w_3=\w_1+\w_2$ or $\w_3=\w_1+\w_2+ 1$ for later use. In the spectral flow conserving case, this reads
\begin{align}
    \langle V_{j_1}^{\w_1}(y_1)V_{j_2}^{\w_2}(y_2)V_{j_3}^{\w_3}(y_3)\rangle \propto & \,\Bigg|\of{1+(-1)^{\w_1} \frac{(\w_3-1)!}{(\w_1-1)!\w_2!}y_3+(-1)^{\w_3}y_2y_3}^{j_1-j_2-j_3}\label{Bosonic3ptFunction}\\
    &\of{1+(-1)^{\w_1+1} \frac{(\w_3-1)!}{(\w_2-1)!\w_1!}y_3+(-1)^{\w_3}y_1y_3}^{j_2-j_1-j_3}\nn\\
    &\of{\frac{\w_3!}{\w_1!\w_2!}-(y_1-(-1)^{\w_3+\w_1}y_2)}^{j_3-j_1-j_2}\Bigg|^2\nn \, ,
\end{align}
up to $y$-independent factors. When spectral flow conservation is violated by one unit, one gets 
\begin{align}
\label{eq: general odd edge}
    &\braket{
    V^{\w_1}_{j_1}(y_1)
    V^{\w_2}_{j_2}(y_2)
    V^{\w_3}_{j_3}(y_3)  } \propto \Bigg|y_3^{j_1 + j_2 - j_3 - \frac{k}{2}} \\
    &\qquad \times 
    \left(1 + (-1)^{\w_1+1} \frac{(\w_1 + \w_2)!}{\w_1! \w_2!}  y_3 + (-1)^{\w_1}y_1 y_3 + (-1)^{\w_3}y_2 y_3 \right)^{\frac{k}{2} - j_1 - j_2 - j_3}\Bigg|^2 \, . \nn 
\end{align}

As will be described below, in the supersymmetric theory the picture changing procedure leads to the necessity of computing some correlators involving affine descendants which are non-trivial in the presence of spectrally flowed insertions \cite{Pakman:2007hn,Giribet:2007wp,Iguri:2022pbp,Iguri:2023khc}. More explicitly, we will be interested in observables of the form  
\begin{align}
   \langle\of{jV}_{j}^{\w}(x,z)\prod_{i}^{n-1}V^{\w_i}_{j_i,h_i}\rangle \,, \qquad 
    \of{jV}_{j}^{\w} \equiv \of{j_{-1-\w}V^{\w}_{j,j+\frac{k}{2}\w}}. 
\end{align}
It was shown in \cite{Iguri:2023khc} that the series identification for lowest-weight states \eqref{series-identif} imply that 
\begin{equation}
   \of{jV}_{j}^{\w} (x,z)=-\N(j)\lim_{y\to\infty }y^{k-2j}D^{-}_{y,\frac{k}{2}-j}V^{\w+1}_{\frac{k}{2}-j}(x,y,z),
\end{equation}
where  $D^{-}_{y,j}= y^{2}\partial_y + 2j y$. It follows that 
\begin{align}
   \langle\of{jV}_{j}^{\w}(x,z)\prod_{i}^{n-1}V^{\w_i}_{j_i,h_i}\rangle = -\N(j)\lim_{y\to\infty }y^{k-2j}D^{-}_{y,\frac{k}{2}-j}\langle V^{\w+1}_{\frac{k}{2}-j}(x,y,z)\prod_{i}^{n-1}V^{\w_i}_{j_i,h_i}\rangle.
\end{align}
It follows that these correlation functions with current insertions can be expressed in terms of the primary ones, albeit with different spectral flow charges and spins.

\subsubsection{Four-point functions}

We now move to four-point functions, and start with the unflowed case.   
Using the global Ward identities gives 
\begin{align}
\begin{aligned}
   &\langle V_{j_1}(x_1,z_1)V_{j_2}(x_2,z_2)V_{j_3}(x_3,z_3)V_{j_4}(x_4,z_4)\rangle = \\[1ex]
   &\qquad \qquad \Bigg|\frac{x_{12}^{-j_1-j_2+j_3-j_4}x_{13}^{-j_1+j_2-j_3+j_4}x_{23}^{j_1-j_2-j_3+j_4}x_{34}^{-2j_4}}{z_{12}^{\Delta_1+\Delta_2-\Delta_3+\Delta_4}z_{13}^{\Delta_1-\Delta_2+\Delta_3-\Delta_4}z_{23}^{-\Delta_1+\Delta_2+\Delta_3-\Delta_4}z_{34}^{2\Delta_4}} \Bigg|^2\mathcal{F}^{0}_{\jj}(x,z) \,,
   \label{unflowed4pt}
\end{aligned}
\end{align}
where $z_{ij}= z_i-z_j$ and $x_{ij}= x_i-x_j$, while 
\begin{equation}
   {\Ff}^{0}_{\jj}(x,z) \equiv \langle V_{j_1}(0,0)V_{j_2}(1,1)V_{j_3}(\infty,\infty)V_{j_4}(x,z)\rangle \,, \qquad\jj=(j_1,j_2,j_3,j_4)\, ,
\end{equation}
with \begin{equation}
    z = \frac{z_{32}z_{14}}{z_{12}z_{34}}\qqquad x = \frac{x_{32}x_{14}}{x_{12}x_{34}} \, . 
\end{equation}
For states in the continuous representations (which are tachyonic for $\w=0$), the OPE reads \cite{Teschner:1999ug,Maldacena:2001km}
\begin{equation}
\label{unflowed OPE}
V_{j_1}(x_1,z_1)V_{j_4}(x_4,z_4)  = \int_{\frac{1}{2} + i \R}dj \int d^2x \frac{\big|z_{14}^{\Delta_j-\Delta_1-\Delta_4}\big|^2C(j_1,j_4,j) V_{1-j}(x,z_1)}{\big|x_{14}^{j_1+j_4-j}(x_4-x)^{j_4+j-j_1}(x-x_1)^{j+j_1-j_4}\big|^2} + \cdots\,, 
\end{equation}
up to descendant contributions weighted by higher powers of $z_{14}$.  Analyzing the short string sector, where $j_1$ and/or $j_4$ take real values, requires a delicate analytic continuation of this expression, which  acquires additional contributions coming from poles of $C(j_1,j_4,j)$ or of the factors $x_{ij}$ crossing the integration contour \cite{Maldacena:2001km}.
By inserting the OPE \eqref{unflowed OPE} into \eqref{unflowed4pt} one obtains the conformal block decomposition 
\begin{equation}           
      {\Ff}^{0}_{\jj}(x,z) = \int_{\frac{1}{2}+i\RR}dj \, {\cal{C}}(j)|F_{j}(x,z)|^2\qqquad
   {\cal{C}}(j)= \frac{C(j_1,j_4,j)C(j,j_2,j_3)}{B(j)} \,.\label{unflowedblocks}
\end{equation} 
The function $F_j(x,z)$ admits an expansion in powers of $z$, namely  
\begin{equation}
\label{unflowedzexp}
    F_j(x,z) = z^{\Delta_{j}-\Delta_{1}-\Delta_{4}} \sum_{n=0}^{\infty} f_{j,n}(x)z^{n} \, ,
\end{equation}
where, due to the Knizhnik-Zamolodchikov (KZ) equation\footnote{There are actually two independent solutions to the KZ equation which are related by the reflection $j\to1-j$, but one can focus on a single one by taking the $j$-integral to run over $\frac{1}{2}+i\RR$ instead of $\frac{1}{2}+i\RR_{+}$.},
\begin{equation}
f_{j,0}(x) = x^{j-j_1-j_4}\pFq{2}{1}{j-j_1+j_4,j-j_3+j_2}{2j}{x} \, .
    \label{unflowed 2F1}
\end{equation}
Higher orders $f_{j,n>0}(x)$ can be obtained recursively \cite{Teschner:1999ug}, although no closed-form expression is known for $F_j(x,z)$. Eqs.~\eqref{unflowed OPE} and \eqref{unflowedblocks} are only valid if the external spins satisfy 
\begin{equation}
{\rm Max}\left[|\Re(j_1-j_4)|,|\Re(1-j_1-j_4)|,|\Re(j_2-j_3)|,|\Re(1-j_2-j_3)|\right]<\frac{1}{2} \,\label{unflowedrange}.
\end{equation}
Away from this range, the correlator picks up additional contributions coming from the exchange of unflowed discrete states and, as shown in  \cite{Maldacena:2001km}, also singly-flowed continuous states. 
Unflowed correlators also satisfy a number of so-called \textit{flip identities} \cite{Dei:2021yom,Dei:2022pkr}, such as 
\begin{align}
    &\langle V_{j_1}(0,0)V_{j_2}(1,1)V_{j_3}(\infty,\infty)V_{j_4}(x,z)\rangle =\nn\\
    &\hspace{2cm}\mathcal{N}(j_1)\mathcal{N}(j_3) |x|^{-4j_4}|z|^{2j_4}\langle V_{\frac{k}{2}-j_1}(0,0)V_{j_2}(1,1)V_{\frac{k}{2}-j_3}(\infty,\infty)V_{j_4}(z/x,z)\rangle\label{flipid} \, .
\end{align}

Spectrally flowed four-point functions are notoriously complicated.  Global Ward identities imply that, in the $y$-basis, 
\begin{align}
&\langle V_{j_1}^{\w_1}(x_1,y_1,z_1)V_{j_2}^{\w_2}(x_2,y_2,z_2)V_{j_3}^{\w_3}(x_3,y_3,z_3)V_{j_4}^{\w_4}(x_4,y_4,z_4) \rangle \nonumber\\[1ex]
& = \Bigg|\frac{ x_{21}^{-h_1^0-h_2^0+h_3^0-h_4^0} x_{31}^{-h_1^0+h_2^0-h_3^0+h_4^0}  x_{32}^{h_1^0-h_2^0-h_3^0+h_4^0} x_{34}^{-2 h_4^0}}
{z_{21}^{-\Delta_1^0-\Delta_2^0+\Delta_3^0-\Delta_4^0} z_{31}^{-\Delta_1^0+\Delta_2^0-\Delta_3^0+\Delta_4^0}  z_{32}^{\Delta_1^0-\Delta_2^0-\Delta_3^0+\Delta_4^0} z_{34}^{-2 \Delta_4^0}} \Bigg|^2\nonumber\\
   &\qquad\!\times \Bigg \langle V_{j_1}^{\w_1} \left(0,\frac{y_1 \, x_{32} \, z_{21}^{\w_1} \, z_{31}^{\w_1} }{x_{21} \, x_{31} \, z_{32}^{\w_1}},0\right) V_{j_2}^{\w_2} \left(1,\frac{y_2 \, x_{31} \, z_{21}^{\w_2} \, z_{32}^{\w_2}}{x_{21} \, x_{32} \, z_{31}^{\w_2}},1\right) V_{j_3}^{\w_3}\left(\infty,\frac{y_3 \, x_{21} \, z_{31}^{\w_3} \, z_{32}^{\w_3}}{x_{31} \, x_{32} \, z_{21}^{\w_3}},\infty\right)\nonumber\\
   &\qquad\qquad\qquad\qquad V_{j_4}^{\w_4} \left(\frac{x_{32} \, x_{14}}{x_{12} \, x_{34}},\frac{y_4 \, x_{31} \, x_{32} \, z_{21}^{\w_4} \,  z_{34}^{2 \w_4}}{x_{21} \, x_{34}^2 \, z_{31}^{\w_4} \,  z_{32}^{\w_4}}, \frac{z_{32} \, z_{14}}{z_{12} \, z_{34}}\right)\Bigg\rangle\ .\label{eq:global Ward identities solution 4pt function}
\end{align}
An integral expression linking these correlators to the unflowed ones was proposed in \cite{Dei:2021yom}. 
It was proposed in \cite{Dei:2021yom} the local Ward identites and KZ equations for 
\begin{equation}
    \langle V_{j_1}^{\w_1}(0,y_1,0)V_{j_2}^{\w_2}(1,y_2,1)V_{j_3}^{\w_3}(\infty,y_3,\infty)V_{j_4}^{\w_4}(x,y_4,z) \rangle \, 
\end{equation}
are solved by
\begin{align}
\begin{aligned}
     &   \big|X_\emptyset^{j_1+j_2+j_3+j_4-k}X_{12}^{-j_1-j_2+j_3-j_4} X_{13}^{-j_1+j_2-j_3+j_4}X_{23}^{j_1-j_2-j_3+j_4}X_{34}^{-2j_4}\big|^2\\[1ex]
    & \quad\quad \times \langle V_{j_1}(0,0)V_{j_2}(1,1)V_{j_3}(\infty,\infty)V_{j_4}\of{\frac{X_{14}X_{23}}{X_{12}X_{34}},z} \rangle\label{even4pt}
\end{aligned}
\end{align}
when the total spectral flow charge is even, and by 
\begin{align}
\begin{aligned}
   &\big| 
   X_{123}^{\frac{k}{2}-j_1-j_2-j_3-j_4} X_{1}^{-j_1+j_2+j_3+j_4-\frac{k}{2}}X_{2}^{j_1-j_2+j_3+j_4-\frac{k}{2}}  X_{3}^{j_1+j_2-j_3+j_4-\frac{k}{2}}
    X_{4}^{-2j_4}\big|^2\\[1ex]
    &\quad \quad  \times
    \langle  \mathcal{N}(j_3)V_{j_1}(0,0)V_{j_2}(1,1)V_{\frac{k}{2}-j_3}(\infty,\infty)V_{j_4}\of{\frac{X_{2}X_{134}}{X_{123}X_{4}},z} \rangle\label{odd4pt}
\end{aligned}
\end{align}
 when it is odd. In order to obtain $x$-basis correlators
\begin{equation}
    \mathcal{F}_{\jj}^{\ww}(x,z) \equiv \langle V^{\w_1}_{j_1h_1}(0,0)V^{\w_2}_{j_2h_2}(1,1)V^{\w_3}_{j_3h_3}(\infty,\infty)V^{\w_4}_{j_4h_4}(x,z)\rangle
\end{equation}
one must then carry out the integral transform
\begin{equation}
    \mathcal{F}_{\jj}^{\ww}(x,z) = 
    \hspace{-0.15cm} \int \prod_{i=1}^{4}d^2y_i \,  \big|y_i^{j_i+\frac{k}{2}\w_i-h_i-1}\big|^2
   \langle V_{j_1}^{\w_1}(0,y_1,0)V_{j_2}^{\w_2}(1,y_2,1)V_{j_3}^{\w_3}(\infty,y_3,\infty)V_{j_4}^{\w_4}(X,y_4,z) \rangle \, ,
\end{equation}
with 
\begin{equation}
    X =  \frac{X_{14}X_{23}}{X_{12}X_{34}} \qquad  \mathrm{or} \qquad 
    X =  \frac{X_{2}X_{134}}{X_{123}X_{4}}\, ,
\end{equation}
depending on whether the total spectral flow is even or odd, respectively. In these expressions, the $X_I$ are the appropriate generalizations of the factors appearing in the three-point functions discussed above, namely the $Z_I$ defined in \eqref{Z_I-3pt}, and $X$ is the generalized cross-ratio. We have 
\begin{equation}
    X_{I} = z^{\frac{1}{2}\delta_{\{1,4\}\in I}}\off{(z-1)(-1)^{\w_1(\w_2+\w_3)+\w_4(\w_2+\w_3)}}^{\frac{1}{2}\delta_{\{2,4\}\in I}} 
    \hspace{-0.2cm}\sum_{i\in I: \vep_i = \pm}\hspace{-0.2cm}P_{\boldsymbol{\w}+\sum_{i\in I}\vep_i \hat{e}_i}\prod_{i\in I} y_i^{\frac{1-\vep_i}{2}} \,,
    \label{XIFunctions}
\end{equation}
where 
\begin{align}
    P_{\boldsymbol{\w}} =& f(\boldsymbol{\w})\of{1-x}^{\frac{1}{2}s(\w_2+\w_4-\w_1-\w_3)}\of{1-z}^{\frac{1}{4}s((\w_1+\w_2-\w_3-\w_4)(\w_1+\w_4-\w_2-\w_3))-\frac{1}{2}\w_2\w_4}\nn\\
    &\times x^{\frac{1}{2}s(\w_1+\w_4-\w_2-\w_3)}z^{\frac{1}{4}s((\w_1+\w_2-\w_3-\w_4)(\w_2+\w_4-\w_1-\w_3))-\frac{1}{2}\w_1\w_4}\tilde{P}_{\boldsymbol{\w}}(x,z)
    \,, \label{Pwfunction} 
\end{align}
with $s(\w) = \w \,\Theta(\w)$, $\Theta$ being the Heaviside step function. Here, the polynomials $\tilde{P}_{\boldsymbol{\w}}(x,z)$ are obtained from the different holomorphic covering maps relevant for a given four-point function, see \cite{Pakman:2009zz,Dei:2021yom}. A necessary condition for the existence of a such maps $\Gamma(z)$ with the appropriate branch points is 
\begin{equation}
    \sum_{i=1}^4 \w_i \, \in \, \mathbb{Z} \, , \quad \text{and}
    \quad \sum_{i=1}^4 \w_i > 2 \,{\rm max} \, \w_i \, .
\end{equation}
This is not a sufficient condition:  once we further impose $\Gamma(z_i)=x_i$ for $i=1,2,3$ there is no freedom left. It follows that the maps $\Gamma(z)$ only exist on the locus where $x_4$ and $z_4$ -- $x$ and $z$ after fixing the other insertion points at $0$, $1$ and $\infty$ -- are appropriately related. The $\tilde{P}_{\boldsymbol{\w}}(x,z)$ are irreducible polynomials which vanish whenever this happens, i.e.~$\tilde{P}_{\boldsymbol{\w}}(x,z) = \prod (z-\Gamma^{-1}(x))$, where the product runs over all possible covering maps and preimages. In particular, they are polynomials of degree $H_{\boldsymbol{\w}}$ in $z$, 
\begin{equation}
    H_{\boldsymbol{\w}} = \frac{1}{2}\min_{i=1,2,3,4}\off{\w_i\of{\sum_{j=1}^4 \w_j - 2\w_i}} \label{Hurwitz}\, 
\end{equation}
being the Hurwitz number, and of degree $\Lambda_{\boldsymbol{\w}}$ in $x$, where   
\begin{equation}
    \Lambda_{\boldsymbol{\w}} = \frac{1}{2}\off{\min(\w_1+\w_2,\w_3+\w_4)-\max(|\w_1-\w_2|,|\w_3-\w_4|)} \,. \label{lambda}
\end{equation}
For more details, see \cite{Dei:2021yom,Dei:2022pkr}. This proposal is supported by several non-trivial checks performed in \cite{Iguri:2024yhb}, where its consistency with the factorization in terms of worldsheet conformal blocks was proved.

\subsection{Supersymmetric vertex operators and holographic dictionary}
We now review the definition of the relevant vertex operators in the supersymmetric theory, i.e.~those  holographically dual to the NSNS chiral primaries of the boundary CFT. In the conventions of \cite{Iguri:2023khc}, omitting the antiholomorphic fermions, and working in the canonical $(-1)$ ghost picture, these operators take the form
\begin{subequations}
\bea
&&\cV^{\omega}_{j}(x,u,z) = \frac{1}{\sqrt{n_5}}e^{-\varphi(z)} \psi^{\omega}(x,z) V^{\omega}_{j}(x,z) 
\chi^{\omega-1}(u,z) W^{\omega}_{j-1}(u,z) \label{Vvertex},\\
&&\cW^{\omega}_{j}(x,u,z) = \frac{1}{\sqrt{n_5}}e^{-\varphi(z)} \psi^{\omega-1}(x,z) V^{\omega}_{j}(x,z) 
\chi^{\omega}(u,z) W^{\omega}_{j-1}(u,z) .\label{Wvertex}
\eea 
\end{subequations}
Here $u$ is the SU$(2)$ isospin variable, $z$ is the worldsheet coordinate, and $x$ the holographic one. We have also introduced the shorthands
\begin{equation}
V^{\omega}_{j}(x,z) \equiv V^{\omega}_{j,j+\frac{k}{2}\w}(x,z) \,, \quad 
W^{\omega}_{l}(u,z) \equiv W^{\omega}_{l,-l+\frac{k'}{2}\w}(u,z)
\label{defVjw(x)}
\end{equation}
such that the spacetime weights are
\begin{equation}
    H \left[\cV^{\omega}_{j}\right] =  j-1+n_5 \w/2 \qqquad H \left[\cW^{\omega}_{j}\right] = 
    j+n_5 \w/2 \,.
    \label{hdefVWw}
\end{equation}
Each vertex operator admits a ghost picture $(0)$ representation 
\begin{equation}
    \Vv^{\omega,(0)}_{j}(x,u,z) = \lim_{w\rightarrow z}\of{e^{\varphi(w)} G(w)} \cV^{\omega}_{j}(x,u,z)\,.
\end{equation}
and similarly for $\cW^{\omega,(0)}_{j}(x,u,z)$, giving 
\begin{equation}
 \Vv^{\w,(0)}_j(x,u,z) = \Aa^{\w,1}_j(x,u,z) + (-1)^\w \Aa^{\w,2}_j(x,u,z) \,,
    \label{Vpicture0}
\end{equation}
with 
\begin{align}
   &\Aa^{\w,1}_j(x,u,z) =\off{j^{-}_{-1-\w}(x,z)- (h-1)\hat{\jmath}^{-}_{-1-\w}(x,z)} \hat{\psi}^\w(x,z) V^\w_j(x,z) \hat{\chi}^{\w}(u,z) W^\w_{j-1}(u,z),\nn \\[1ex]
   &\Aa^{\w,2}_j(x,u,z) =-\frac{1}{n_5}\off{k^{+}_{\w}(u,z)-(h-1)\hat{k}^+_w(u,z)}\psi^{\w}(x,z)V^{\w}_j(x,z)\chi^{\w}(u,z)W^\w_{j-1}(u,z) \nn,
\end{align}
and
\begin{equation}
    \Ww^{\w,(0)}_j(x,u,z) = \Bb^{\w,1}_j(x,u,z) + (-1)^\w \Bb^{\w,2}_j(x,u,z) \,,
        \label{Wpicture0}
\end{equation}
with 
\begin{align}
    &\Bb^{\w,1}_j(x,u,z) =\off{k^{-}_{-1-\w}(u,z)-h\,\hat{k}^{-}_{-1-\w}(u,z)} \hat{\psi}^\w(x,z) V^\w_{j}(x,z)\hat{\chi}^{\w} (u,z) W^\w_{j-1}(u,z) ,\nn \\[1ex]
   &\Bb^{\w,2}_j(x,u,z) =\frac{1}{n_5} \off{j^{+}_{\w}(x,z) + h\,\hat{\jmath}^{+}_\w(x,z)} \psi^{\w}(x,z)V^\w_j(x,z)\chi^\w(u,z)W^\w_{j-1}(u,z) \nn, 
\end{align}
where $h=j + \frac{n_5}{2} \w$. 
For the purposes of this work, it will be useful to work with chiral operators for which the SU$(2)$ primary is taken to have lowest-weight, i.e.~to be of the form $W_{l,-l}^{\w}(z)$. This is done by taking
\begin{align}
&\cV^{\omega}_{j}(x,z) \equiv  \cV^{\omega}_{j}(x,u,z)\Big|_{u=0} \qqquad \cW^{\omega}_{j}(x,z) \equiv  \Ww^{\omega}_{j}(x,u,z)\Big|_{u=0} .
\end{align}
Anti-chiral operators are defined analogously from $W_{l,l}^{\w}(z)$, giving  
\begin{align}
&\cV^{\dagger\omega}_{j}(x,z) \equiv  \cV^{\omega}_{j}(x,u,z)\Big|_{u=\infty} \qqquad \cW^{\dagger\omega}_{j}(x,z) \equiv  \Ww^{\omega}_{j}(x,u,z)\Big|_{u=\infty} .
\end{align}

We now describe these operators from the point of view of the boundary theory at the symmetric orbifold point, namely the Sym$^{N}\left(T^4\right)$ CFT with $N=n_1 n_5$  \cite{Argurio:2000tb,Dabholkar:2007ey,Giribet:2007wp}. At large $N$, the holographic dictionary maps  single string states in the bulk to single cycle fields of this HCFT. The latter are constructed from the bare twist fields, conventionally denoted $\sigma_n$, which must be suitably dressed. In the $T^4$ case, each twist sector gives rise to four types of chiral primary operators \cite{Lunin:2001pw}, which we respectively label $O_n^-(x)$, $O_n^+(x)$, and $O_n^a(x)$ with $a=1,2$. Their holomorphic weights are 
\begin{equation}
    H \left[O_n^{-} \right] = \frac{n-1}{2}  \,,\quad
    H \left[O_n^{a} \right] = \frac{n}{2} \,,\quad
    H \left[O_n^{+} \right] = \frac{n+1}{2} \,;\quad
 n=1,2,\dots.    \label{D1D5CFTweights3}
\end{equation}
As per AdS/CFT, these are associated with $z$-integrated $x$-basis operators of the worldsheet theory. Thus, up to the normalization (to be discussed below), for the NSNS sector the identification is as given by 
\begin{equation}
    O_n^- (x,\bar{x}) \,  \sim \,   
    \int d^2z \, \Vv^{\omega}_{j}(x,\bar{x},z,\bar{z}) \qqquad O_n^+ (x,\bar{x}) \, \sim \,   \int d^2z \, \Ww^{\omega}_{j}(x,\bar{x},z,\bar{z}) ,  
\end{equation}
together with  
\begin{equation}
 n \,=\, 2 j - 1 + n_5 \, \w \,. \label{njw}
\end{equation}
From the worldsheet point of view, the allowed ranges are 
\begin{equation}
    j = 1,\frac{3}{2},\dots,
\frac{n_5}{2}
,\qquad
\w = 0,1,\dots \, .
\label{jwrangecc}
\end{equation}
The worldsheet theory is shown to describe all chiral primaries of the holographic CFT, except for those in twisted sectors where $n$ is a non-zero multiple of $n_5$ 
\cite{Seiberg:1999xz,Dabholkar:2007ey,Giribet:2007wp}
see however \cite{Eberhardt:2018vho}. The precise normalization 
\begin{equation}
    \mathbb{O}^{\omega}_{j}(x,\bar{x},z,\bar{z}) = \frac{\Oo^{\omega}_{j}(x,\bar{x},z,\bar{z})}{\sqrt{2c_\nu^{-1}n_5\sigma^2\of{2h-1}B(j)v_4}}\qqquad\sigma=\sqrt{\frac{b^2\gamma(-b^2)}{4\pi \nu }}.
    \label{ONormNS}
\end{equation}
was obtained in \cite{Iguri:2022pbp}, where $\Oo$ ($\mathbb{O}$) stands for either $\Vv$ ($\mathbb{V}$) or $\Ww$ ($\mathbb{W}$).

Correlators involving single-cycle operators are computed by means of a lift from the base space to the appropriate covering surface. It turns out that, in the four-point case, the condition 
\begin{equation}
    n_3 = n_1+n_2+n_4-2\label{Sphere condition}
\end{equation}
on the twists $n_i$ (up to permutations) guarantees that the genus zero contribution is the only non-trivial one. Among the extremal four-point functions computed in \cite{Pakman:2009ab}, the one relevant for the present work is
\begin{align}
\label{extremal holographic 1}
&\<O_{n_1}^{-}(0)O_{n_2}^{-}(1)O_{n_3}^{-^{\dagger}}(\infty)O_{n_4}^-(x,\xb)\> = F_4(n_i)\frac{n_3^{3}}{(n_1n_2n_3n_4)^{1/2}}\,, 
\end{align}
where 
\begin{equation}
\label{extremal holographic 2}
    F_4(n_i)=\off{\frac{(N-n_1)!(N-n_2)!(N-n_4)!}{(N-n_3)!(N!)^2}}^{1/2} \to \frac{1}{N} \, . 
\end{equation}
Here we have included the large $N$ limit of $F_4(n_i)$ in the final expression.
As it was proved in \cite{Pakman:2009ab}, double-cycle exchanges provide an account for part of the leading order contribution to extremal correlators even in the large $N$ limit. The combined effect of both single- and double-cycle operators was shown to coincide with the single-cycle contribution multiplied by a factor $n_3/\tilde{n}$, with $\tilde{n}=n_1+n_4-1$. 
Therefore, we can isolate the single-cycle (sc) contribution as 
\begin{equation}
\label{extremal holographic single-cycle}
    \<O_{n_1}^{-}(0)O_{n_2}^{-}(1)O_{n_3}^{-^{\dagger}}(\infty)O_{n_4}^-(x,\xb)\>_{\text{sc}} = F_4(n_i)\frac{n_3^{2}(n_1+n_4-1)}{(n_1n_2n_3n_4)^{1/2}}\,,
\end{equation}
Additionally, there exists a family of non-extremal that also  satisfy \eqref{Sphere condition}. These read 
\begin{align}
&\<O_n^{+}(0)O_2^{-}(1)O_{n+2}^{-\dagger}(\infty)O_2^{-\dagger}(x,\xb)\> = G(x,\xb)\,.
\end{align}
At small \( x \) we have 
\begin{equation}
G(x,\xb) \approx \frac{(n+2)^{3/2}}{2(n+1)n^{1/2}} \sqrt{\frac{(N-n)(N-n-1)}{N^2(N-1)^2}} |x|^{-2} \to \frac{1}{N}\frac{(n+2)^{3/2}}{2(n+1)n^{1/2}}  |x|^{-2} \, ,
\label{holographic non extremal}
\end{equation}
where in the last step we have taken the large $N$ limit.

\section{Spacetime factorization of the bosonic four-point function}\label{Sec: On the bosonic four-point function}

Before delving into the supersymmetric case, it will be fruitful to study the spacetime OPEs of the bosonic AdS$_3$ model. Here we remain agnostic about the internal sector of the geometry AdS$_3\times X$, where $X$ is some a compact manifold with the correct dimensionality needed to construct the critical bosonic string background. 
Given that, in addition, for the supersymmetric case we will mainly be interested in extremal four-point functions, we focus on bosonic correlators of the form 
\begin{equation}
    \Ff^{\ww}_{\jj}(x,z) =\langle V_{j_1}^{\w_1}(0,0)V_{j_2}^{\w_2}(1,1)V_{j_3}^{\w_3}(\infty,\infty)V_{j_4}^{\w_4}(x,z) \rangle, \label{LWBosonic4pt} 
\end{equation}
which satisfy\footnote{These are the spectral flow \textit{conserving} four-point functions in the original notation of \cite{Cagnacci:2015pka}.}
\begin{equation}
    \w_3 = \w_1+\w_2+\w_4\, , \label{W3condition}
\end{equation}
and where all vertex operators are taken to be the spectrally flowed image of a lowest weight SL$(2,\RR)$ primary, namely  
\begin{equation}
    V_{j}^\w(x,z) \equiv V^\w_{j,h=j+\frac{k}2{\w}}(x,z) = \lim_{y\to 0} V_j^\w(x,y, z) \, .
\end{equation}
In Eq.~\eqref{LWBosonic4pt} we have also used the shorthands $\jj$ and $\ww$, representing the relevant set of unflowed spins $j_i$ and spectral flow charges $\w_i$. 

\subsection{Setting up the stage}

Let us discuss the general features of the correlator \eqref{LWBosonic4pt}. By using the integral expression of  \cite{Dei:2021yom}, it can be expressed as 
\begin{align}
\hspace{-0.17cm}    \Ff^{\ww}_{\jj}(x,z) =
    \Big|X_{\emptyset}^{j_1+j_2+j_3+j_4-k}X_{12}^{-j_1-j_2+j_3-j_4}X_{13}^{-j_1+j_2-j_3+j_4}X_{23}^{j_1-j_2-j_3+j_4}X_{34}^{-2j_4}\Big|^2\Ff^{0}_{\jj}\of{X,z}. 
    \label{Eberhardt-HW}
\end{align}
Here, $X$ is the generalized cross-ratio, and all factors $X_I$, defined in Eq.~\eqref{XIFunctions},  should be evaluated at $y_i=0$.  Furthermore,  whenever the constraint \eqref{W3condition} is satisfied, the Hurwitz number $H_{\ww}$ defined in \eqref{Hurwitz} vanishes, hence several of the relevant polynomials $P_{\ww} (x,z)$ trivialize. As a result, apart from simple powers of $z$ and $z-1$, the only $X_I$ that remain different from $1$ are 
\begin{align}
    & X_{14} = f_{14}\tilde{P}_{14}(x,z)z^{w_1\w_4-\frac{(\w_1+1)(\w_4+1)-1}{2}}(1-z)^{s[\w_4(\w_2-1)]-\frac{(\w_4+1)\w_2}{2}}x^{s(1-\w_2)}\,,\\
    &X_{12}= f_{12}\tilde{P}_{12}(x,z)z^{s[\w_1(\w_4-1)]-\frac{(\w_1+1)\w_4}{2}}(1-z)^{s[\w_2(\w_4-1)]-\frac{\w_4(\w_2+1)}{2}}\,, 
\end{align}
which leads to 
\begin{align}
    X =\frac{X_{23}X_{14}}{X_{12}X_{34}}= \frac{f_{14}f_{23}}{f_{12}f_{34}}\frac{x^{s(1-\w_2)}(1-z)^{\w_4\delta_{\w_2,0}+\w_2\delta_{\w_4,0}}}{z^{\w_1\delta_{w_4,0}}}\frac{\tilde{P}_{14}(x,z)}{\tilde{P}_{12}(x,z)}\label{Crossratio}
    \, .
\end{align}
The polynomials $\tilde{P}_{14}(x,z)$ and $\tilde{P}_{12}(x,z)$ are degree 1 in $x$ and degree $\w_3$ in $z$.  The terms that will be relevant for our analysis take the form 
\begin{align}
        f_{14}\tilde{P}_{14}&(x,z)= \label{P14 exp}\\
        &z^{\w_3} + \dots + Q_{(\w_3-\w_2+2,\w_3,\w_2)}z^{\w_3-\w_2+1} +x\off{Q_{(w_1+1,w_3-w_2,\w_4+1)}+\dots + \xi_{14} z^{w_4+1}}\nn \, ,\\
        f_{12}\tilde{P}_{12}&(x,z)= \label{P12 exp}\\
        &z^{\w_3} + \dots + Q_{(\w_3-\w_2+1,\w_3,\w_2+1)}z^{\w_3-\w_2} +x\off{Q_{(w_1+1,w_3-w_2-1,\w_4)}+\dots + \xi_{12} z^{w_4}} \, . \nn
\end{align}
Importantly, here the $Q_{\ww}$ are the coefficients appearing in spectrally flowed three-point functions  \eqref{3pt-final}.

The spacetime four-point function is given by the worldsheet integral
\begin{equation}
 \Ff_{\text{target}}(x) = \int d^2z \Ff_{\jj}^{\ww}(x,z)\Ff_{\text{int}}(z), 
 \label{spacetime bosonic}
\end{equation}
where $\Ff_{\text{int}}(z)$ accounts for the internal contribution, and the factorization limit corresponds to studying how Eq.~\eqref{spacetime bosonic} behaves at small $x$. Although the full integral is presently out of reach, here we will show that, even in the spectrally flowed case, Eq.~\eqref{Eberhardt-HW} allows us to study this correlation function along the lines of what was done in \cite{Maldacena:2001km} for the unflowed four-point function.
For this, we separate the worldsheet integral into 
\begin{align}
    \text{region I:}\quad |z|<1 \qqquad  \text{region II:}\quad |z|>1,
\end{align}
and center our discussion on the contribution from region I. The contribution from region II is expected to be irrelevant for single-particle exchanges \cite{Maldacena:2001km}.

\begin{figure}[h!]
    \centering
    \includegraphics[scale=0.15]{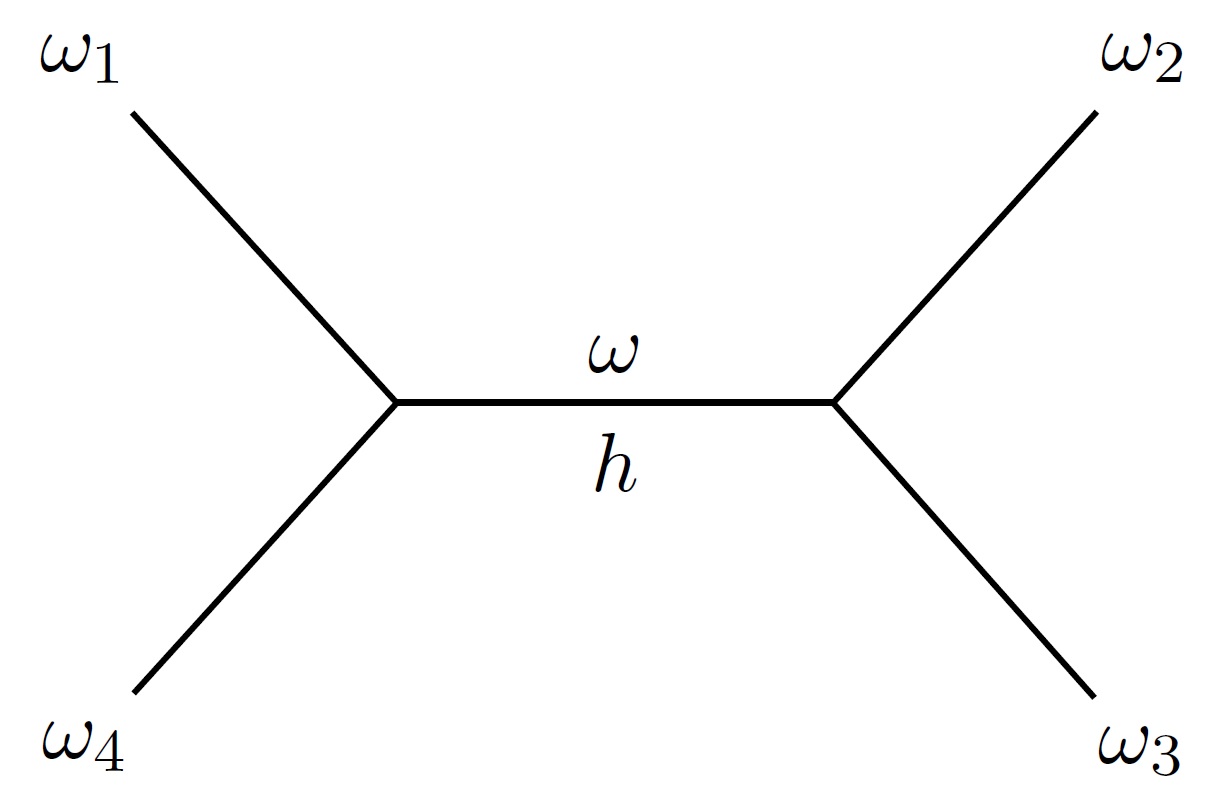}
    \caption{Schematic description of the factorization along the $14\to23$ channel.}
    \label{fig: Basic diagram}
\end{figure}

We now look for a change of variables that allows us to zoom in on the small $z$ region as we take $x\to 0$ without losing relevant information. More precisely, the goal is to keep the generalized cross-ratio $X$ fixed in the $x\to 0$ limit. In order to understand how to do this, it is  useful to discuss what the possible exchanged states are.  As it was discussed for the unflowed case in \cite{Maldacena:2001km} and more generally in \cite{Dei:2021xgh,Bufalini:2022toj}, even though all external states belong to the discrete representations of SL(2,$\R$), the theory allows for the exchange of both short and long strings. A given intermediate state with spectral flow charge $\w$ must satisfy the selection rules associated with each of the relevant three-point functions, namely \cite{Maldacena:2001km}
\begin{equation}
    \w_i+\w_j \geq \w_k-1 \qquad \forall \,\, 
    i \neq j \neq k \, . 
    \label{fusionrules}
\end{equation}
In the region $|z|<1$ (and since we work at small $x$), these correspond to the \textit{left} and \textit{right} three point functions for the $s$-channel exchange of Fig.~\ref{fig: Basic diagram}.
Assuming that \eqref{W3condition} holds, this immediately implies that $\w$ is restricted to the range 
\begin{equation}
    \w_1 + \w_4 + 1 \geq \w \geq \w_3-\w_2-1 = \w_1+\w_4-1 \, , 
\end{equation}
so there seem to be only three possibilities. Actually, for the case at hand we can do a little better. Since all external states with spectral flow charges $\w_i$ are built upon lowest weight operators, they can also be interpreted as states with $\w_i' = \w_i+1$ by means of the series identifications in Eq.~\eqref{series-identif}. Making use of this for the operator with the highest charge, namely $V_{j_3}^{\w_3}$, we conclude that 
\begin{equation}
    \w_1 + \w_4 + 1 \geq \w \geq \w_3'-\w_2-1 = \w_1+\w_4 \, , 
    \label{w range bos}
\end{equation}
In other words, only channels with $\w=\w_1 + \w_4 =  \w_3-\w_2$ or $\w=\w_1 + \w_4 + 1 =  \w_3-\w_2+1$ are allowed. As will become clear shortly, it is not a coincidence that these are precisely the smaller powers of $z$ appearing in the $x$-independent terms of 
Eqs.~\eqref{P14 exp} and \eqref{P12 exp}. This will play a key role in defining the appropriate limit for the generalized cross-ratio $X$ appearing in the flowed four-point function. 

\subsection{The $\w=\w_1+\w_4$ channel}

We first consider the exchange of states with $
    \w=\w_1+\w_4 = \w_3-\w_2$ 
since this will be the most relevant one for the supersymmetric case, where the external states are chiral primaries. 
This can be analyzed by defining a new integration variable
\begin{equation}
    u = \frac{x}{z^\w} \label{u w change}\, . 
\end{equation}
Region I thus corresponds to $|x|< |u|$. We find that, in the small $x$ limit with fixed $u$, the worldsheet four-point function \eqref{LWBosonic4pt} reduces to
\begin{align*}
    \Ff^{\ww}_{\jj}(x,z) &= \Big|z^{-j_4 \w_1 -j_1\w_4-k\frac{\w_1\w_4}{2}}\off{Q_{l}\of{a+b_4 u}}^{-j_1-j_2+j_3-j_4}
    \Big|^2\Ff^{0}_{\jj}\off{\frac{u}{a+b_4 u},z},
\end{align*}
where $z=z(u)$ and 
\begin{align}
    Q_{l} = \dfrac{\w!}{\w_1!\w_4!},\quad a = \dfrac{\w_1!\w_4!\w_3!}{\w_2!(\w!)^2},\quad b_4 = \dfrac{\w_4}{\w} \label{Qlab4}.
\end{align}
Here the last factor on the right corresponds to the unflowed four-point function, whose conformal block can be expanded as in Eqs.~\eqref{unflowedzexp} and \eqref{unflowed 2F1}. 
On the other hand, the internal conformal block can be expanded as usual, giving 
\begin{equation}
    \Ff_{\text{int}}(z) \sim \sum_{\Delta'} \Big|z^{\Delta'-\Delta'_1-\Delta'_4}\Big|^2 C_{\text{int}}(\Delta')\,,
\end{equation}
with $C_{\text{int}}(\Delta')$ the relevant product of three-point functions (assuming the internal two-point functions are unit-normalized). Hence, we find that, at leading order, the contribution from Region I takes the form 
\begin{align}
    \Ff_{\text{target}}(x)\big|_{|z|<1}&= \int_{|x|<|u|}d^2u \int_{\frac{1}{2}+i\RR} dj \sum_{\Delta'} C_{\text{int}}(\Delta') \Cc(j) \, \Big|Q_{l}^{-j_1-j_2+j_3-j_4} \w^{-1}  
    \label{Ftarget bos} \\
    &\hspace{-0.5cm} \times z^{\alpha_z}u^{j-j_1-j_4-1}\of{a+b_4 u}^{j_3-j-j_2}
    \pFq{2}{1}{j-j_1+j_4,j-j_3+j_2}{2j}{\frac{u}{a+b_4 u}} \nn\Big|^2\,,
\end{align}
where the overall power of $z$ is 
\begin{align}
    \alpha_z
    =\Delta+\Delta'-(j_1+j_4)\w- \frac{k}{4}\w^2-\Delta^{\w_1}_1-\Delta'_1-\Delta^{\w_4}_4-\Delta'_4+1 \, .
\end{align}
Here the external SL$(2,\RR)$ weights read $\Delta^{\w}_i=\Delta_i - j_i \w_i - k \w_i^2/4$, hence the corresponding Virasoro conditions read $\Delta^{\w_i}_i+\Delta'_i=1$. Since we are dealing with operators built upon lowest-weight states, for the external spacetime weights we have $h_i = j_i + \frac{k}{2}\w_i$. As for the intermediate state, assuming it has $\w=\w_1+\w_4$, the on-shell condition will imply that its spacetime weight $h = m  + \frac{k}{2}\w$ must satisfy 
\begin{equation}
    h  \w = \
    \Delta+\frac{k}{4}\w^2+\Delta'-1 \, . 
\end{equation}
Consequently, we can write    
\begin{align}
    \alpha_z=\w \of{h- \frac{k}{2}\w-j_1-j_4} = \w\of{h- h_1-h_4} \, . 
\end{align}
Though the spacetime cross-ratio $x$ does not appear explicitly in Eq.~\eqref{Ftarget bos}\footnote{There is actually an additional dependence on $x$ in the integration limit for $u$. We will show below that this does not contribute.}, it is contained in $z$, which we think of as a function of $u$, namely $z^{\w}= x/u$. This shows that the overall power of $x$ is of the form  $x^{h-h_1-h_4}$,  precisely what is expected from the   OPE expansion in the boundary theory. 

From the discussion above, we conclude that 
\begin{align}
    \Ff_{\text{target}}(x)\big|_{|z|<1}&= \int_{|x|<|u|}d^2u \int_{\frac{1}{2}+i\RR} dj \sum_{\Delta'} C_{\text{int}}(\Delta')\Cc(j) \Big|Q_{l}^{-j_1-j_2+j_3-j_4} x^{h-h_1-h_4} \w^{-1} \label{Bosonic4ptSpacetime} \\
    &\times u^{j-h+\frac{k}{2}\w-1}\of{a+b_4 u}^{j_3-j-j_2}
    \pFq{2}{1}{j-j_1+j_4,j-j_3+j_2}{2j}{\frac{u}{a+b_4 u}}\Big|^2\,. \nn
\end{align}
We now deal with the integration over the complex variable $u$. As in \cite{Maldacena:2001km}, we need to regularize it in order to be allowed to perform it  \textbf{before} integrating over the intermediate unflowed spin $j$. This is done by first integrating over the whole $u$-plane and then subtracting the integrals over the region $|u|<|x|$, such that
\begin{align}
     \Ff_{\text{target}}(x)\big|_{|z|<1} =  \Ff^{\CC}_{\text{target}}(x)-\Ff^{|x|}_{\text{target}}(x) 
     \label{u regions} \, .
\end{align}
In principle, one should also include potential contributions from the regions near $a+b_4 u=0$ and $u\to\infty$, although we expect that they are irrelevant for single-particle exchange. We now discuss each of the contributions appearing on the RHS of \eqref{u regions} separately. 

\subsubsection{Long string exchange}

Let us first focus on the contribution from the full complex plane, where the relevant integral reads  
\begin{align}
    &\Ii_{\CC}= \int_\CC d^2u \Big|u^{j-h+\frac{k\w}{2}-1}\of{a+b_4 u}^{j_3-j-j_2}
    \pFq{2}{1}{j-j_1+j_4,j-j_3+j_2}{2j}{\frac{u}{a+b_4 u}}\Big|^2 \, .
\end{align}
 By introducing $b_1 = \w_1/\w$,
 rescaling $u\to au$, and using the identity \eqref{HypIden}, this can be expressed as 
\begin{align}
    \Ii_{\CC}=\frac{|a^{j_3-j_2-h+\frac{k\w}{2}}|^2\pi\gamma(2j)}{\gamma(j+j_2-j_3)\gamma(j+j_3-j_2)} I \, , 
    \end{align}
    where, inside the $j$-integral, one can take  
\begin{equation}
    I= \int d^2u  \, d^2y \Big|u^{j-h+\frac{k\w}{2}-1}y^{j+j_2-j_3-1}(1-y)^{j+j_3-j_2-1}(1-b_1 u y)^{j_1-j-j_4}(1+b_4 u y)^{j_4-j-j_1}\Big|^2 \, .\nn
\end{equation} 
Importantly, this double complex integral decouples after rescaling $y\to y/u$, and then rescaling $u\to y u$, giving 
\begin{align}
    I= &\int d^2u \big|u^{-j-h+\frac{k\ww}{2}}(1-u)^{j_3+j-j_2-1} \big|^2 \int d^2y  \big| y^{j-h+\frac{k\w}{2}-1}(1-b_1y)^{j_1-j_4-j}(1+b_4y)^{j_4-j_1-j}\Big|^2 .\nn
\end{align}
Then, by taking $y\to Q_l y$ and $u\to Q_r u$, where $Q_l$ was defined in Eq.~\eqref{Qlab4} and $Q_r= \w! \w_2!/\w_3!$, we find that
\begin{align}
    &\Ff^{\CC}_{\text{target}}(x) = \sum_{\Delta'} C_{\text{int}}(\Delta')\int_{\frac{1}{2}+i\RR}dj  |x|^{2(h-h_1-h_4)} \nn \\
    & \qquad \times \frac{C(j,j_2,j_3) \pi\gamma(2j)Q_{r}^{2(j_2-j_3-j+1)}}{B(j)\gamma(j+j_2-j_3)\gamma(j-j_2+j_3)} \int d^2u \Big|u^{-j-h+\frac{k\ww}{2}}(1-Q_{r} u)^{j_3+j-j_2-1}\Big|^2  \\ 
    & \qquad \times C(j_1,j_4,j)Q_{l}^{2(j-j_1-j_4)}\int d^2y \Big|y^{j-h+\frac{k\w}{2}-1}(1-Q_{l} b_1y)^{j_1-j_4-j}(1+Q_{l} b_4y)^{j_4-j_1-j}\Big|^2 \nn\, .
\end{align}
This should match the product of relevant $y$-basis three-point functions, for which the general formula was provided in  Eq.~\eqref{Bosonic3ptFunction}, further divided by the corresponding propagator. Indeed, this is exactly what happens: by making use of the reflection symmetry \eqref{Creflection}, our result can be written as 
\begin{align}
    &\Ff^{\CC}_{\text{target}}(x)  \label{final long bos} \\
    &\qquad = \sum_{\Delta'} C_{\text{int}}(\Delta') \int_{\frac{1}{2}+i\RR}dj  |x|^{2(h-h_1-h_4)} \langle V^{\w_1}_{j_1}(0)V^{\w_4}_{j_4}(1)V^{\w}_{j,h}(\infty)\rangle \langle V^{\w}_{1-j,h}(0)V^{\w_2}_{j_2}(1)V^{\w_3}_{j_3}(\infty)\rangle \nn\\
    & \qquad = \sum_{\Delta'} C_{\text{int}}(\Delta')\int_{\frac{1}{2}+i\RR}dj  |x|^{2(h-h_1-h_4)} \frac{\langle V^{\w_1}_{j_1}(0)V^{\w_4}_{j_4}(1)V^{\w}_{j,h}(\infty)\rangle \langle V^{\w}_{j,h}(0)V^{\w_2}_{j_2}(1)V^{\w_3}_{j_3 }(\infty)\rangle}{\langle V^{\w}_{j,h}(0)V^{\w}_{j,h}(\infty)\rangle} \, .
    \nn
\end{align}
We thus identify this contribution as describing the exchange of long string states. 

\subsubsection{Short string exchange}

We should stress that in this derivation, and in contrast with the discussion of \cite{Iguri:2024yhb}, the spacetime weight $h$ of the intermediate state is not fixed by the external ones. It can be thought of as a function of the remaining quantum numbers of the intermediate state, i.e.~the unflowed spin $j$ and the spectral flow charge $\w$, defined through the Virasoro condition of the worldsheet theory.

We now move to the integral of $u$ over the inner circle $|u|<|x|$. For small $x$, we can safely expand the integrand as a power series, hence the relevant integral reads 
\begin{align}
    \Ii_{|x|} 
    &=\int_{|u|<|x|} d^2 u \Big|a^{j_3-j_2-h+\frac{k}{2}\w}\sum_{p=0}^\infty u^{j+p-h+\frac{k}{2}\w-1} \label{Ix bos}\\
    &\quad \times \frac{(j-j_3+j_2)_{p}}{(2j)_{p}p!}\sum_{n=0}^p(j-j_1+j_4)_{n}\binom{p}{n}(2j+n)_{p-n} (-b_4)^{p-n}\Big|^2\, .\nn
\end{align}
After reinserting the anti-holomorphic contributions, we have 
\begin{equation}
    \int_{|u|<|x|}d^2u |u|^{2(j-h+\frac{k}{2}\w-1)}u^{p}\bar{u}^{\bar{p}} =\frac{\pi |x|^{2(j+p+\frac{k}{2}\w-h)}\delta_{p,\bar{p}}}{j+p+\frac{k}{2}\w-h}.
\end{equation}
This shows that, for each term in the sum over $p$ one gets an additional pole for the $j$-integral localized at 
\begin{equation}
    h = j+p+\frac{k}{2}\w \, . 
    \label{h short bos}
\end{equation}
This ensures that the overall power of $x$ is unchanged, and  shows that such contributions describe the exchange of short string states belonging to a flowed lowest-weight representation. Here $h = h(j)$ should be understood as  depending implicitly on $j$ due to the Virasoro condition
\begin{equation}
 \Delta -h\w+\frac{k}{4}\w^2+\Delta'=1. \label{intermediate Virasoro}
\end{equation}
Hence, for each value of $p$ the resulting residue is computed from  \cite{Cardona:2010qf}
\begin{equation}
    \int dj \frac{|x|^{2\lambda(j)} g(j)}{\lambda(j) } =  \frac{2\pi g(j)}{\partial_j[\la(j)]} \Bigg|_{j=j_0} \, ,\qquad \lambda(j_0)=j_0+p+\frac{k}{2}\w-h(j_0) = 0 \, . \label{j-pole integral}
\end{equation}
for a given analytic function $g(j)$ which stands for the remaining factors in the original integral \eqref{Ix bos}. 
Consequently, the present contribution to the spacetime correlator gives a sum over factors of the form
\begin{align}
    &\Ff^{|x|,p}_{\text{target}}(x)=\frac{2\pi^2}{n_5(2j+n_5\w-1)}|x|^{2(h-h_1-h_4)}\sum_{\Delta'} C_{\text{int}}(\Delta')
    \label{final short bos} \\
    &\times \off{\sum_{n=0}^p\Big|\frac{1}{p!}\binom{p}{n}(j-j_1+j_4)_{n}(2j+n)_{p-n} (-b_4)^{p-n}\Big|^2}
    \frac{\off{C(j,j_2,j_3)(j-j_3+j_2)_p}}{\off{B(j)(2j)_p}}.\nn
\end{align}
Here the overall factor coming from the integration over $j$ gives the correct  normalization factor for the intermediate state. Keeping it aside, we find that, for each value of $p$, the second line in Eq.~\eqref{final short bos} is nothing but the relevant coefficient appearing in the power series $y$-basis expansion for the relevant product of three-point functions, see  Eq.~\eqref{final long bos}. These coefficients are, by definition,  precisely the products of spacetime three-point functions involving short-string states.

\subsection{The $\w=\w_1+\w_4+1$ channel}

As discussed around Eq.~\eqref{w range bos}, we expect to find one (and only one) additional contribution to the factorization of our four-point function: one for which spectral flow is not conserved at each of the three-point functions involved. This channel  can also be isolated by changing variables to $
u = \frac{x}{z^\w}$, now with $\w = \w_1+\w_4+1$. The computation follows analogously to the spectral flow conserving case, and the main difference resides in the behavior of the generalized cross-ratio $X$. 

Let us briefly describe how this works. As before, we concentrate on the small $z$ region. From Eq.~\eqref{Crossratio} we see that, at leading order in $x$ and at fixed $u$, the generalized cross-ratio takes the form
\begin{align}
    X = z \hspace{1mm} \frac{Q_{(\w+1,\w_3,\w_2)} + Q_{(\w_1+1,\w-1,\w_4+1)}u}{Q_{(\w+1,\w_2+1,\w_3)}}, 
\end{align}
Due to the overall dependence on $z$, it is convenient to  use a flip identity \eqref{flipid} at the level of the unflowed four-point function. This leads to a contribution from the SL(2,$\R$) sector which reads 
\begin{align}
    \Ff^{\ww}_{\jj}&(x,z) =\Nn(j_1)\Nn(j_3)\Big|z^{-\w_4j_1-(\w_1+1)j_4-\frac{k}{2}\w_1\w_4}{Q^{j_3-j_1-j_2-j_4}_{(\w+1,\w_2+1,\w_3)}}\\&\of{{Q_{(\w+1,\w_3,\w_2)} + Q_{(\w_1+1,\w-1,\w_4+1)}u}}^{-2j_4}\Big|^2\tilde\Ff^{0}_{\jj}\off{\frac{Q_{(\w+1,\w_2+1,\w_3)}}{Q_{(\w+1,\w_3,\w_2)} + Q_{(\w_1+1,\w-1,\w_4+1)}u},z}, \nn
\end{align}
where $\tilde{\Ff}^0_\jj$ has the unflowed spins $j_{1}$ and $j_3$ replaced by $\frac{k}{2}-j_{1}$ and $\frac{k}{2}-j_{3}$, respectively. Inserting the relevant conformal block expansion (at leading order in $z$)
\begin{align}
    \tilde\Ff^{0}_{\jj}(X,z)= \int_{\frac{1}{2}+i\RR}dj \tilde{\Cc}(j)\Big|z^{\Delta-\Delta_1-\Delta_4-j_1+\frac{k}{4}}X^{j+j_1-j_4-\frac{k}{2}}\tilde{F}\off{X}\Big|^2\,,
\end{align}
with 
\begin{align}
    \tilde{F}\off{X}=\pFq{2}{1}{j+j_1+j_4-\frac{k}{2},j+j_2+j_3-\frac{k}{2}}{2j}{X},
\end{align}
and
\begin{align}
    \tilde\Cc(j)=\frac{C\of{\frac{k}{2}-j_1,j_4,j}C\of{j,j_2,\frac{k}{2}-j_3}}{B(j)},
\end{align}
and further including the internal sector as well as the $u$-integral as before, we obtain 
\begin{align}
    \Ff^{\ww}_{\jj}(x,z) =& \int_{|x|<|u|}d^2u\int_{\frac{1}{2}+i\RR}dj \sum_{\Delta'}\Nn(j_1)\Nn(j_3)\tilde\Cc(j)\Cc'(\Delta')\Big|{Q^{j+j_3-j_2-\frac{k}{2}}_{(\w+1,\w_2+1,\w_3)}} \w^{-1}
    \\
    &\times x^{h-h_1-h_4}u^{h_1+h_4-h-1}\of{{Q_{(\w+1,\w_3,\w_2)} + Q_{(\w_1+1,\w-1,\w_4+1)}u}}^{\frac{k}{2}-j-j_1-j_4}\nn\\&\times\tilde{F}\off{\frac{Q_{(\w+1,\w_2+1,\w_3)}}{Q_{(\w+1,\w_3,\w_2)} + Q_{(\w_1+1,\w-1,\w_4+1)}u}}\Big|^2\,.\nn
\end{align}
Here, we have introduced the appropriate spacetime conformal weight $h = \frac{\Delta+\Delta'-1}{\w}+\frac{k}{4}\w$ of the intermediate state consistent with a spectral flow charge $\w=\w_1+\w_4+1$ and with the worldsheet on-shell condition \eqref{intermediate Virasoro}.

In contrast with the previous discussion, for this channel  the expected \textit{left} and \textit{right} three-point functions are parity-odd functions in terms of their total spectral flow. It is worth noticing that the usage of the flip identity immediately introduces the necessary factors $\Nn(j_{1})$ and $\Nn(j_{3})$, which, together with the function $\tilde\Cc(j)$, precisely match the overall normalization  of the related three-point functions, see Eq.~\eqref{consdei}. 

For the sake of brevity we focus on the contributions coming from the exchange of long strings. As before, this corresponds to the piece of the regularized integral where $u$ is integrated over the full complex plane. After rescaling $u\to \frac{Q_{(\w+1,\w_3,\w_2)}}{Q_{(\w_1+1,\w-1,\w_4+1)}}u$, the relevant integral reads
\begin{align}
    \tilde{\Ii}_\CC=\Big|\frac{Q_{(\w+1,\w_3,\w_2)}^{-j-h+\frac{k}{2}\w}}{Q^{h_1+h_4-h}_{(\w_1+1,\w-1,\w_4+1)}}\Big|^2\int_{\CC}d^2u \Big|u^{h_1+h_4-h-1}\of{{1-u}}^{\frac{k}{2}-j-j_1-j_4}\tilde{F}\off{\frac{a}{1 - u}}\Big|^2\,,\nn
\end{align}
where $a=Q_{(\w+1,\w_2+1,\w_3)}/Q_{(\w+1,\w_3,\w_2)}$. Parity-odd three-point functions are somewhat simpler, hence the resulting hypergeometric integral can be carried out explicitly. By taking $t = 1/(1-u)$ and integrating over $t$ we get 
\begin{align}
\begin{aligned}
    \tilde{\Ii}_\CC &=\frac{\gamma(h_1+h_4-h)\gamma(j+h-\frac{k}{2}\w)}{\gamma(j+j_1+j_4-\frac{k}{2})} \\ 
    & \quad \times \Big|\frac{Q_{(\w+1,\w_3,\w_2)}^{-j-h+\frac{k}{2}\w}}{Q^{h_1+h_4-h}_{(\w_1+1,\w-1,\w_4+1)}}\pFq{2}{1}{j+j_2+j_3-\frac{k}{2},j+h-\frac{k}{2}\w}{2j}{a}\Big|^2\,.
\end{aligned}
\end{align}
Using Eq.~\eqref{eq: general odd edge}, it is now easy to show that 
\begin{equation}
    \Ff^{\ww}_{\jj}(x,z) = \int_{\frac{1}{2}+i\RR}dj \sum_{\Delta'}\Cc'(\Delta')|x|^{2(h-h_1-h_4)}\langle V^{\w_1}_{j_1,h_1}V^{\w_4}_{j_4,h_4}V^{\w_1+\w_4+1}_{j,h}\rangle\langle V^{\w_1+\w_4+1}_{1-j,h}V^{\w_2}_{j_2,h_2}V^{\w_3}_{j_3,h_3}\rangle\,,
\end{equation}
which is the expected factorized expression due to the reflection identity \eqref{reflection w>0}. Processes involving the exchange of short strings can be described following a procedure analogous to what was done in previous sections.

This shows explicitly that, as expected from the selection rules derived in \cite{Maldacena:2001km} and further discussed in \cite{Dei:2021xgh,Bufalini:2022toj}, the factorization of four-point functions in the SL(2,$\R$) model involves channels which do not conserve spectral flow. This is true even when the four-point function itself does not describe a spectral flow violating process. Our analysis constitutes a  generalization of the original analysis of \cite{Maldacena:2001km}, where the authors showed that a consistent factorization of the boundary four-point function with discrete external states with $\w_i=0$ necessarily involves the exchange of long strings with $\w=1$. Here we have shown that this holds for generic values of $\w_i$ with $i=1,2,3,4$, although we have focused on the case with $\w_3=\w_1+\w_2+\w_4$ for simplicity.

Let us finish this section by summarizing what we have achieved. We have considered a spacetime four-point function of operators which, in the worldsheet language, were spectrally flowed lowest-weight states satisfying $\w_3 = \w_1+\w_2+\w_4$, and described how to isolate all relevant contributions to its factorization. This involves the exchanges of both long and short strings, belonging to the two channels allowed by the spectral flow selection rules: $\w = \w_1+\w_4$ and $\w=\w_1+\w_4+1$. In all cases, we have started from the worldsheet description for strings in AdS$_3$ (while remaining agnostic about the compact CFT associated with the internal sector) and derived the $x$-dependence required by consistency with the OPE of the holographic CFT. We have also shown that one obtains the expected normalization, i.e.~the relevant product of three-point functions. Indeed, once the insertion points in $z$ and $x$ are fixed to $(0,1,\infty)$ and the relative normalization is taken into account, these take the same form in the boundary and worldsheet CFTs. We now extend our analysis to the supersymmetric case, where four-point functions that are protected by supersymmetry can be compared explicitly with those obtained from the symmetric orbifold CFT. 


\section{Extremal four-point functions of chiral primary operators}\label{sec: Extremal four-point functions of chiral primary operators}
Throughout this section we  focus on the four-point function for strings in AdS$_3\times S^3 \times T^4$ involving only chiral primary operators polarized along the extended directions, namely 
\begin{equation}
    \langle \Vv_{j_1}^{\w_1,(0)}(0,0)\Vv_{j_2}^{\w_2}(1,1)\Vv_{j_3}^{\dagger\w_3,(0)}(\infty,\infty)\Vv_{j_4}^{\w_4}(x,z)\rangle \, .\label{SUSY correlator}
\end{equation}
The vertex operators $\Vv^{\w_i}_{j_i}(x,z)$ were defined in \eqref{Vvertex}, see also \eqref{Vpicture0} for the corresponding picture zero operators. In analogy with what happened when computing chiral three-point functions \cite{Iguri:2022pbp,Iguri:2023khc}, fermion number counting implies that only a single term from each picture zero operator contributes non-trivially. Here and in the following formulas we  omit the antiholomorphic sector. The supersymmetric correlator thus takes the form 
\begin{align}
    &\mathbb{F}_\jj^{\ww}(x,z) = \langle e^{-\varphi(1)}e^{-\varphi(z)}\rangle\langle W^{\w_1}_{j_1-1}W^{\w_2}_{j_2-1}W^{\dagger\w_3}_{j_3-1}W^{\w_4}_{j_4-1}\rangle\langle\hat{\chi}^{\w_1}\hat{\chi}^{\w_2}\hat{\chi}^{\w_3}\hat{\chi}^{\w_4}\rangle\\
    &\langle \off{\of{j^{-}- (h_1-1)\hat{\jmath}^{-}}_{-1-\w_1} \hat{\psi}^{\w_1} V^{\w_1}_{j_1}}\psi^{\w_2}V^{\w_2}_{j_2}\off{\of{j^{-}- (h_3-1)\hat{\jmath}^{-}}_{-1-\w_3} \hat{\psi}^{\w_3} V^{\w_3}_{j_3}}\psi^{\w_4}V^{\w_4}_{j_4}\rangle\,.\nn
\end{align}
The worldsheet insertions are taken to be $(z_1,z_2,z_3,z_4)=(0,1,\infty,z)$, while for the SL(2,$\R$) sector we also take $(x_1,x_2,x_3,x_4)=(0,1,\infty,x)$. It is convenient to divide the computation into four parts, according to the different SL(2,$\R$) current insertions, namely
\begin{align}
    \mathbb{F}_\jj^{\ww}(x,z) = (h_3-1)\mathbb{F}^{\ww}_1(x,z)+(h_1-1)\mathbb{F}^{\ww}_3(x,z)+(h_1-1)(h_3-1)\mathbb{F}^{\ww}_0(x,z)+\mathbb{F}^{\ww}_{13}(x,z), \nn
\end{align}
where  
\begin{align}
\begin{aligned}
    &\mathbb{F}_0^{\ww}(x,z) = \mathsf{W}^{\w}_{\jj}
    \langle \of{\hat{\jmath}^{-}_{-1-\w_1} \hat{\psi}^{\w_1}}\psi^{\w_2}\of{\hat{\jmath}^{-}_{-1-\w_3} \hat{\psi}^{\w_3}}\psi^{\w_4}\rangle\langle V^{\w_1}_{j_1}V^{\w_2}_{j_2} V^{\w_3}_{j_3}V^{\w_4}_{j_4}\rangle\,,\\
    &\mathbb{F}_1^{\ww}(x,z) = \mathsf{W}^{\w}_{\jj}
    \langle \hat{\psi}^{\w_1}\psi^{\w_2}\of{\hat{\jmath}^{-}_{-1-\w_3} \hat{\psi}^{\w_3}}\psi^{\w_4}\rangle\langle\of{j^{-}_{-1-\w_1}V^{\w_1}_{j_1}}V^{\w_2}_{j_2} V^{\w_3}_{j_3}V^{\w_4}_{j_4}\rangle\,,\\
    &\mathbb{F}_3^{\ww}(x,z) = \mathsf{W}^{\w}_{\jj}
    \langle \of{\hat{\jmath}^{-}_{-1-\w_1} \hat{\psi}^{\w_1}}\psi^{\w_2}\hat{\psi}^{\w_3}\psi^{\w_4}\rangle\langle V^{\w_1}_{j_1}V^{\w_2}_{j_2} \of{j^{-}_{-1-\w_3}V^{\w_3}_{j_3}}V^{\w_4}_{j_4}\rangle\,,\\
    &\mathbb{F}_{13}^{\ww}(x,z) = \mathsf{W}^{\w}_{\jj}
    \langle \hat{\psi}^{\w_1}\psi^{\w_2}\hat{\psi}^{\w_3}\psi^{\w_4}\rangle\langle\of{j^{-}_{-1-\w_1}V^{\w_1}_{j_1}}V^{\w_2}_{j_2}\of{j^{-}_{-1-\w_3}V^{\w_3}_{j_3}}V^{\w_4}_{j_4}\rangle\,,
    \label{F0F1F3F13}
\end{aligned}
\end{align}
with 
\begin{equation}
    \mathsf{W}^{\ww}_{\jj}=\langle e^{-\varphi(1)}e^{-\varphi(z)}\rangle\langle W^{\w_1}_{j_1-1}(0)W^{\w_2}_{j_2-1}(1)W^{\dagger\w_3}_{j_3-1}(\infty)W^{\w_4}_{j_4-1}(z)\rangle\langle\hat{\chi}^{\w_1}(0)\hat{\chi}^{\w_2}(1)\hat{\chi}^{\w_3}(\infty)\hat{\chi}^{\w_4}(z)\rangle. \nn
\end{equation}
As in the bosonic case, the spacetime correlator $\mathbb{F}^{\ww}_{\text{target}}(x)$ is obtained by integrating over $z$, and we  focus on the contribution from the Region I: $|z|<1$. Following  \cite{Maldacena:2001km}, we expect the $|z|>1$ to be relevant only for two-particle exchanges.  

Let us first discuss the bosonic SU(2) correlator. In terms of the so-called isospin variable \cite{Zamolodchikov:1986bd} (which plays a role analogous to the $x$ variable in SL(2,$\R$)) the four-point function is, in the language of \cite{Dei:2021xgh,Bufalini:2022toj}, in the so-called coincidence limit. The explicit computation is carried out in Appendix \ref{sec: App B - The SU(2) sector}. This also holds for the  correlator involving the SU(2) fermions, which corresponds to the particular case with $k'=2$ and $l_i=0$, since operators $\hat{\chi}^\w$ are nothing but spectrally flowed fermionic identities. Combining these two factors gives 
\begin{align}
\mathsf{W}_\jj^{\ww}=\frac{|z|^{2(\boldsymbol{\Delta}'^{\w}(l)-\boldsymbol{\Delta}'^{\w_1}_1-\boldsymbol{\Delta}'^{\w_4}_4)}}{|1-z|^2}\delta_{j_3,j_1+j_2+j_4-2}\cC'(j_i-1,l)\, , \label{SU2 and ghost}
\end{align}
where we have included the product of unflowed SU(2) structure constants, 
\begin{equation}
    \cC'(l_i,l) = C'(l_1,l_4,l)C'(l,l_2,l_3) \, ,
\end{equation}
and with $\boldsymbol{\Delta}'^{\w_i}_i = \Delta'_i+\w l_i +(k'+2) w^2/4$. In principle on should include a sum over the SU(2) spin $l$ of the intermediate state. However, charge conservation for the full four-point under consideration implies the extremality condition 
\begin{equation}
        j_3=j_1+j_2+j_4-2. \label{ExtremalJcondition}
\end{equation}
Given that the external states satisfy $l_i = j_i-1$, this implies $l_3-l_2=l_1+l_4$. Combined with the well-known SU$(2)$ selection rules, we find that the sum over $l$ collapses to a single term, namely the contribution with $l=j_3-j_2=j_1+j_4-2$. 

The SL(2,$\R$) sector is more involved, in particular because one must deal with correlators that include current insertions -- which are non-trivial in the presence of spectrally flowed operators -- such as 
\begin{align}
   (\Jj_1\Ff)^{\ww}_{\jj}(x,z)&= \langle (j^{-}_{-1-\w_1}V^{\w_1}_{j_1})(0)V^{\w_2}_{j_2}(1)V^{\w_3}_{j_3}(\infty)V^{\w_4}_{j_4}(x,z)\rangle, \\
   (\Jj_3\Ff)^{\ww}_{\jj}(x,z)&= \langle V^{\w_1}_{j_1}(0)V^{\w_2}_{j_2}(1)(j^{-}_{-1-\w_3}V^{\w_3}_{j_3})(\infty)V^{\w_4}_{j_4}(x,z)\rangle, \\
   (\Jj_1\Jj_3\Ff)^{\ww}_{\jj}(x,z)&= \langle (j^{-}_{-1-\w_1}V^{\w_1}_{j_1})(0)V^{\w_2}_{j_2}(1)(j^{-}_{-1-\w_3}V^{\w_3}_{j_3})(\infty)V^{\w_4}_{j_4}(x,z)\rangle, 
\end{align}
together with their fermionic counterparts. These four-point functions can be computed following the method outlined in \cite{Iguri:2023khc}, which we now summarize. In the $y$-basis, the action of each current can be represented in terms of a differential operator, giving  
\begin{equation}
   (j^{-}_{-1-\w}V^{\w}_{j})(x,z)=-\N(j)\lim_{y\to\infty }y^{k-2j}\left[y^{2}\partial_y + (k-2j) y\right]V^{\w+1}_{\frac{k}{2}-j}(x,y,z) \, .
\end{equation}
At the level of correlators, this leads to 
\begin{align}
 & (\Jj_1\Ff)^{\ww}_{\jj}(x,z)= \label{J1FDiff} \\ 
 & \qquad -\N(j_1) \lim_{y_1\to\infty }y_1^{k-2j_1}\left[y^{2}\partial_y + (k-2j) y\right]\langle V^{\w_1+1}_{\frac{k}{2}-j_1}(y_1)V^{\w_2}_{j_2}(y_2)V^{\w_3}_{j_3}(y_3)V^{\w_4}_{j_4}(y_4)\rangle\Big|_{y_{i \neq 1}=0} \, .\nn
\end{align}
The explicit form of the relevant  $y$-basis four-point function  depends both on the spectral flow parity of the original correlator and on the number of current insertions. By means of  Eqs.~\eqref{W3condition},  \eqref{J1FDiff} and \eqref{flipid}, we find that, for the present example, the generalized cross-ratio takes the form 
\begin{equation}
    X_{\text{odd}}= z\frac{X_{123}X_{4}}{X_2X_{134}}.
\end{equation}
This matches the one found in the bosonic computation, written in Eq.~\eqref{Crossratio}, in the limit where we take $y_1\to \infty$ and $y_2,y_3,y_4\to0$. 
A similar discussion holds for the correlators $(\Jj_3\Ff)^{\ww}_{\jj}$, and $(\Jj_1\Jj_3\Ff)^{\ww}_{\jj}$. We thus find that these correlators can  be expressed in terms of the unflowed primary four-point function and its derivatives. 

From now on, we simply follow the guidelines of the bosonic computation discussed in Sec.~\ref{Sec: On the bosonic four-point function}. For the supersymmetric analysis we focus on the the channel with $\w=\w_1+\w_4=\w_3-\w_2$ for simplicity, and change variables to $
    u = x/z^{\w}$. We now discuss in detail (holomorphic contributions to) the contribution from $(\Jj_1\Ff)^{\ww}_{\jj}$. From  \eqref{J1FDiff} one finds that, in the relevant regime,   
\begin{align}
    (\Jj_1\Ff)^{\ww}_{\jj}(x,u)=&Q_{l} u^{1+j_1+j_2-j_3+j_4} (Q_{l}/X)^{-j_1-j_2+j_3-j_4} z^{-j_4 \w_1-(j_1-1)
   \w_4-\frac{1}{2} k \w_1\w_4-1} \\
   &\times \off{X \der_X
    +
  \frac{\of{2j_4 (\w_1+1)+\w_4(k-2j_1)}}{\w_1+1}} \Ff^{0}_{\jj}\off{X(u),z}\nn,
\end{align}
where $X(u)=\frac{u}{a+b_4 u}$, $Q_l, a$ and $b_4$ were defined  in \eqref{Qlab4}, $z$ should be understood as a function of $x$ and $u$, and $\Ff^{0}_{\jj}[X,z]=\langle V_{j_1}(0,0)V_{j_2}(1,1)V_{j_3}(\infty,\infty)V_{j_4}(X,z)\rangle$.
The latter is expressed in terms of the conformal block expansion,  and we keep only the leading term at small $z$, see Eq.~\eqref{unflowedzexp}. After imposing the extremality condition \eqref{ExtremalJcondition}, we get 
\begin{align}
    (\Jj_1\Ff)^{\ww}_{\jj}(x,u)=&\int_{\frac{1}{2}+i\RR} dj \Cc(j)z^{\Delta-\Delta_1-\Delta_4-j_4 \w_1-(j_1-1)
   \w_4-\frac{1}{2} k \w_1\w_4-1}(Q_l u)^{-1} X^{2+j-j_1-j_4}
   \\
   & \times \off{\frac{(j-j_1+j_4)(1+\w_1)+\w_4(k-2j_1)}{1+\w_1}{}_2F_1\off{X}+X 
   {{}_2F_1}'\off{X}} \nn,
\end{align}
where we have introduced the short-hand notation 
\begin{equation}
    {}_2F_1\off{X}\equiv \pFq{2}{1}{j-j_1+j_4,j-j_3+j_2}{2j}{X}\label{HypShortHand}.
\end{equation}
The fermionic SL$(2,\RR)$ sector can be studied analogously after setting $j_2,j_4\to-1$, $j_1,j_3\to0$ and $k\to-2$ \cite{Iguri:2023khc}. More explicitly, each flowed fermionic four-point function with current insertions is related to some unflowed fermionic two-point function, and is therefore completely determined. For instance, for 
\begin{equation} 
    (\hat{\Jj}_3\hat{\Ff})^{\ww}(x,z)= \langle \hat\psi^{\w_1}(0)\psi^{\w_2}(1)(\hat{\jmath}^{-}_{-1-\w_3}\hat\psi^{\w_3})\psi^{\w_4}(x,z)\rangle\,,
\end{equation}
the procedure above simply leads to  
\begin{equation} 
    (\hat{\Jj}_3\hat{\Ff})^{\ww}(x,u)= -2 Q_ln_5^2 (a-b_1u)z^{\w_1(1+\w_4)}.
\end{equation}
Combining the SL(2,$\R$) bosonic and fermionic correlators with their SU(2) counterparts, and further including the Jacobian for the change of variables from $z$ to $u$, we finally obtain 
\begin{align}
\begin{aligned}
    \hspace{-0.25cm}\int_{|z|<1} d^2z \mathbb{F}^{\ww}_1(x,z) =& \int_{|x|<|u|}d^2u \int_{\frac{1}{2}+i\RR}dj \Cc(j) \Cc'(l)\frac{n_5^2}{\w}z^{\alpha_z}u^{-2}X^{j-j_1-j_4-2}(a-b_1u) \label{J1FcorrelatorSUSY}\\
   &\times\off{\frac{(j-j_1+j_4)(1+\w_1)+\w_4(k-2j_1)}{1+\w_1}{}_2F_1\off{X}+X{{}_2F_1}'\off{X}}.
   \end{aligned}
\end{align}
Using the Virasoro conditions for the external states, we get  $\alpha_z=\Delta+\Delta'$. The rest of the contributions in Eq.~\eqref{F0F1F3F13} give similar results
, which can be derived from 
\begin{align}
\begin{aligned}
    (\Jj_3\Ff)^{\ww}_{\jj}(x,z)=&(Q_{l} (a+b_4 u))^{-j_1-j_2+j_3-j_4-1} z^{-j_1\w_4-j_4\w_1
   -\frac{k \w_1\w_4}{2}}\label{J3bosonic} \\
   & \times \off{ X \der_X+\of{j_1+j_2-j_3+j_4}}\Ff^{0}_{\jj}\off{X(u),z}\,,
   \end{aligned}
\end{align}
and
\begin{align}
\begin{aligned}
   (\Jj_1\Jj_3\Ff)^{\ww}_{\jj}(x,z)&=uQ_l(Q_{l} (a+b_4 u))^{-j_1-j_2+j_3-j_4-1} z^{-1-(j_1-1)\w_4-j_4\w_1
   -\frac{k \w_1\w_4}{2}}\label{J3J1bosonic} \\
   &\hspace{-1cm} \Bigg\{ \off{\of{j_1+j_2-j_3+j_4+1}+\frac{2j_4 (\w_1+1)+\w_4(k-2j_1)}{\w_1+1}}X\der_X \\
   &\hspace{-1cm}+\frac{2j_4 (\w_1+1)+\w_4(k-2j_1)}{\w_1+1}(j_1+j_2-j_3+j_4) + X^2\der^2_X \Bigg\}  \Ff^{0}_{\jj}\off{X(u),z}\,.
\end{aligned}
\end{align}

As will become clear below, we are actually describing the exchange of a supersymmetric state polarized along the  AdS$_3$ directions,  described by the  operator
\begin{align}
    \Vv^{\w}_{\text{int}}(x,z)\sim e^{-\varphi(z)}\psi^{\w}(x,z)V^{\w}_{jh}(x,z)\hat\chi^{\w}(z)W^{\w}_{l,-l}(z) \, , \label{Psi-Intermediate}
\end{align}
with $\w=\w_1+\w_4$ and $ l=j_1+j_4-2$. The Virasoro condition for such state reads
\begin{equation}
    \Delta - \left(m-1\right)\w +\frac{n_5}{4}\w^2+ \Delta' + (j_1+j_4-2)\w -\frac{n_5}{4}\w^2=0 \label{Intermediate Virasoro cond}
\end{equation}
where $m=h-\frac{k}{2}\w$ is the unflowed spin projection of the intermediate state. This establishes the relation between the (supersymmetric) spacetime weight $H=m+\frac{n_5}{2}\w-1$ and $\Delta = \Delta(j)$, hence the $x$-dependence in  \eqref{J1FcorrelatorSUSY} is obtained from the overall power of $z$, namely  
\begin{equation}
    z^{\alpha_z}= z^{(m-j_1-j_4+1)\w}= \of{\frac{x}{u}}^{m-j_1-j_4+1} = \of{\frac{x}{u}}^{H-H_1-H_4}\, . 
\end{equation}
Here $H_i = j_i-1+\frac{n_5}{2}\w_i$. As expected, this overall power of $x$ is consistent with the OPE of the supersymmetric boundary theory.

Let us now gather all contributing to the supersymmetric correlator. Upon inverting the order of the $j$ and $u$ integrals by following the regularization  procedure discussed in Sec.~\ref{Sec: On the bosonic four-point function}, including the anti-holomorphic factors, and further changing variables to $X = u/(a+b_4u)$, we get 
\begin{align}
     \mathbb{F}^{\ww}_{\text{target}}(x)=\int_{\frac{1}{2}+i\RR}&dj|x|^{2(m-j_1-j_4+1)}\mathfrak{C}(j,l)n_5^4\int_{\text{reg}}d^2X \Big|X^{j-m-1}(1-b_4 X)^{m-j_1-j_4}\mathfrak{F}[X]\Big|^{2} \, , \label{Reg4pt}
\end{align}
where 
\begin{align}
\begin{aligned}
    \mathfrak{F}[X] =& \, (1-X)^2X^2{}_2F''_1[X]+\Bigg[\of{h_1+h_4+j+n_5\frac{\w}{2}-2}\of{h_2+h_3+j+n_5\frac{\w}{2}-2}
    \\&-2\of{h_4+j+n_5 \frac{\w}{2}-1}\of{h_2+j+n_5 \frac{\w}{2}-1}X \\
    &+\of{h_4-h_1+j+n_5\frac{\w}{2}}\of{h_2-h_3+j+n_5\frac{\w}{2}}X^2\Bigg]{}_2F_1[X] \\
    &+4\off{h_3+j+n_5\frac{\w}{2}-\frac{1}{2}+\of{h_1-j-n_5 \frac{\w}{2}-\frac{3}{2}}X}(1-X)X {}_2F_1'[X] \, , 
    \label{HyperCombination}
\end{aligned}
\end{align}
and 
\begin{align}
    \mathfrak{C}(j,l) = \Cc(j) \Cc'(l)a^{j_1+j_4-m-1}\w^{-1} \, .
    \label{chiral 4pt norm}
\end{align}

Eqs.~\eqref{Reg4pt}-\eqref{chiral 4pt norm} constitute one of our main technical results. In what follows, we will show how to derive from these expressions the relevant information about the different contributions to the spacetime OPE.
We expect three types of terms coming from the different pieces of the regularized integral over $X$, and depending on what propagates along the $s$-channel. In particular, we will obtain long strings exchanges, encoded in the integral over the complex plane, and short strings contributions, which appear as a discrete set of residues from the $j$-integration, and come from the small $X$-regime. We proceed as in the bosonic case, and denote these contributions as 
\begin{align}
     \mathbb{F}^{\ww}_{\text{target}}(x)= \mathbb{F}^{\ww}_{\CC}(x)+\mathbb{F}^{\ww}_{|x|}(x) \, ,
\end{align}
respectively, and treat them separately in the following sections.

\subsection{Exchange of chiral primary states}\label{sec: Exchange of chiral primary states}

We expect the information about the exchange of chiral primary states to emerge from the worldsheet moduli integration over the region $|X|<|x|$. This is precisely what happened in the unflowed case, studied in \cite{Cardona:2010qf}. These contributions are directly related to the structure constants and fusion rules associated with the  chiral ring of the spacetime theory, which are protected by supersymmetry and can thus be compared directly with results obtained from the HCFT at the symmetric orbifold point.

In the small $X$ region, the leading term obtained from Eq.~\eqref{Reg4pt} takes the form 
\begin{align}
   \mathbb{F}^{\ww}_{|x|}(x)&= \int_{\frac{1}{2}+i\RR} dj |x|^{2(m-j_1-j_4+1)}\mathfrak{C}(j,l)n_5^4\\
   &\quad\times|(h_1+h_4+j+\frac{n_5}{2}\w-2)(h_3+h_2+j+\frac{n_5}{2}\w-2)|^2\int_{|X|<|x|} d^2X  |X|^{2(j-m-1)}.\nn
\end{align}, 
where $\w m=\Delta+\Delta'+(l+1)\w$.
Integrating over $X$ in this region gives 
\begin{align}
    \int_{|X|<|x|} d^2X  |X|^{2(j-m-1)} = \frac{\pi}{j-m}|x|^{2(j-m)} \label{X power integral}\, .
\end{align}
As in the bosonic case, the regularization procedure gives rise to an additional pole for the $j$-integral, located at $j = m$. Recall that $m=m(j)$, hence the pole is defined by the condition 
\begin{align}
    0=\lambda(j)=j-m(j)=\frac{j(j-1)}{n_5\w} - \frac{l(l+1)}{n_5\w}+(j-l-1) \, . \label{Virasoro susy from pole}
\end{align}
Here we have used that $l=j_1+j_4 -2$ due to charge conservation. We find that, for an extremal four-point function, the leading channel corresponds to the exchange of a chiral primary state, for which Eq.~\eqref{Virasoro susy from pole} is the Virasoro condition. 

The integration over $j$  then amounts to 
\begin{align}
    \int dj \pi \frac{|x|^{2\lambda(j)}}{\lambda(j)} = \frac{2\pi^2}{\partial_j(\lambda)|_{j_0}} \label{j power integral} \, , 
\end{align}
where $\lambda(j_0)=0$, i.e.~$j_0=l+1=j_1+j_4-1$. As it turns out, at this locus the product of unflowed SL$(2,\RR)$ and SU$(2)$ structure constants drastically simplifies \cite{Dabholkar:2007ey}, and the overall coefficient $\mathfrak{C}(j,l)$ becomes
\begin{equation}
    \mathfrak{C}(j,l) = \mathfrak{C}(j,j-1) = \frac{\sigma^{2}}{\w}\prod_{i=1}^4 \sqrt{B(j_i)},\quad\sigma=\sqrt{\frac{b^2\gamma(-b^2)}{4\pi \nu }},
\end{equation}
where $B(j)$ was defined in Eq.~\eqref{B(j)}. Therefore, at leading order in $x$ we obtain  
\begin{align}
    \mathbb{F}^{\ww}_{|x|}(x) = |(h_1+h_4+h-2)(h_3+h_2+h-2)|^2\frac{2\pi^{2}\sigma^{2}n_5^4}{\w\partial_j(\lambda(j))|_{j_0}}\prod_{i=1}^4 \sqrt{B(j_i)}\, .
\end{align}
Using the extremality condition \eqref{ExtremalJcondition} and $j=j_1+j_4-1$ we can rewrite this as 
\begin{align}
    \mathbb{F}^{\ww}_{|x|}(x) = |(2h-1)(2h_3-1)|^2\frac{2\pi^{2}n_5^4\sigma^{2}}{\w\partial_j(\lambda(j))|_{j_0}}\prod_{i=1}^4 \sqrt{B(j_i)}\, .
\end{align}

We now compare this result with those of the boundary theory, provided in Eqs.~\eqref{extremal holographic 1}, \eqref{extremal holographic 2} and \eqref{extremal holographic single-cycle}, and originally obtained in \cite{Pakman:2009ab}. We first note that, as expected, the leading contribution to the extremal four-point function turns out to be independent of $x$. In order to compare the corresponding coefficient it is necessary to normalize the operators appropriately. The correct normalization for spectrally flowed states, proposed in \cite{Giribet:2007wp} and later derived in \cite{Iguri:2022pbp}, is given in Eq.~\eqref{ONormNS}. 
It is worth noticing that the Jacobian from the $j$-integral, namely  
\begin{align}
    \frac{1}{\w\partial_j(\lambda(j))|_{j_0}}= \frac{n_5^2}{2h-1}\, ,
\end{align}
introduces precisely the (square of the) expected normalization for the intermediate state, as it is necessary for the correct factorization  in terms of the relevant boundary three-point functions. Finally, in order to complete the holographic matching we fix the free parameter of the WZW model, $\nu$,  as \cite{Dabholkar:2007ey,Giribet:2007wp,Iguri:2022pbp}
\begin{equation}
    \nu = \frac{2\pi^5}{b^4\gamma(1+b^2)}, \quad b^2=n_5^{-1} \, . 
\end{equation}
Inserting the string coupling $g_s^2=\frac{n_5}{n_1}v_4$, together with an additional factor of $v_4$, the internal volume, our worldsheet calculation shows that the leading short string contribution to the spacetime correlator gives  
\begin{align}
   \langle \VV_{j_1}^{\w_1}(0)\VV_{j_2}^{\w_2}(1)\VV_{j_3}^{\dagger\w_3}(\infty)\VV_{j_4}^{\w_4}(x)\rangle = \frac{1}{N}\frac{(2h-1)(2h_3-1)^{2}}{\sqrt{(2h_1-1)(2h_2-1)(2h_3-1)(2h_4-1)}}\, ,
\end{align}
where $ N=n_1 n_5$. When written in terms of the twists $n_i$ of the boundary insertions, this exactly reproduces the large $N$ limit of the single-cycle contribution to the four-point function in the holographic CFT, namely Eq.~\eqref{extremal holographic single-cycle} \cite{Pakman:2009ab}. This holds for all extremal correlators. In particular, the spectral flow charges are unconstrained, apart from the condition $\w_3=\w_1+\w_2+\w_4$. Our result thus generalizes the one obtained in \cite{Cardona:2010qf} for unflowed operators to all spectrally flowed sectors of the theory. We further expect extremal correlators involving spacetime chiral primaries that are dual to worldsheet NSNS operators polarized along the SU(2) directions and/or those belonging to the RNS sector of the theory to be computed analogously.

\subsection{Exchange of long string states}

As in the bosonic case, long string channels are expected to arise from integrating $X$ over the full complex plane,  
\begin{align}
     \mathbb{F}^{\ww}_{\CC}(x)=\int_{\frac{1}{2}+i\RR}&dj|x|^{2(m-j_1-j_4+1)}\mathfrak{C}(j,l)n_5^4\int_{\CC}d^2X \Big|X^{j-m-1}(1-b_4 X)^{m-j_1-j_4}\mathfrak{F}[X]\Big|^{2}\,.\label{LongCont}
\end{align}
We now show that this expression can be identified with the appropriate product of three-point functions for a given factorization channel. This corresponds to the exchange of states created by vertex operators of the form\footnote{As usual, here we omit the antiholomorphic dependence in order to simplify the notation. This includes the antiholomorphic variables $\xb$ and $\zb$, and also the left-handed fermions. It is important to keep in mind that the latter can be chosen independently. Hence, to be more precise, one should write $\Vv^{\w}_{r\bar{r}}$ instead of $\Vv^{\w}_{r}$, and, for instance, the RHS of Eq.~\eqref{suma polarizaciones} should be understood as a sum over both $r$ and $\bar{r}$.} 
\begin{align}
    \Vv^{\w}_{r}(x,z)= e^{-\varphi(z)}\psi^{r,\w}(x,z)V^{\w}_{j,m-1-r}(x,z)\hat\chi^{\w}(z)W^{\w}_{l,-l}(z) \, , \label{LongString Vertex}
\end{align}
again with $r=-,0,+$. These operators were argued to be the correct ones for long string states in the supersymmetric model \cite{Yu:2024kxr,Sriprachyakul:2024gyl}, see also \cite{Iguri:2022pbp}.
Note that the different operators defined in \eqref{LongString Vertex} share the same worldsheet and spacetime conformal dimensions. The latter is given by $H=m-1+n_5\w/2$, where 
\begin{equation}
   m = \frac{\Delta+\Delta'}{\w}  + (j_1+j_4-1).
\end{equation}
The factorized expansion we obtain is thus a linear combination of the  form
\begin{align}
\label{suma polarizaciones}
     \mathbb{F}^{\ww}_{\CC}(x)\sim\int_{\frac{1}{2}+i\RR}&dj|x|^{2(m-j_1-j_4+1)}\sum_{r=0,\pm}\langle\Vv^{\w_1,(0)}_{j_1}\Vv^{\w_4}_{j_4} \Vv^{\w}_{j,r}\rangle \langle\Vv^{\w}_{1-j,r} \Vv^{\w_2}_{j_2}\Vv^{\w_3,(0)}_{j_3}\rangle.
\end{align}
Let us first focus on the SL$(2,\RR)$ sector, where the \textit{left} and \textit{right} three point functions involving one long string are given by 
\begin{align}
    &\langle\of{\psi^{\w_1} V^{\w_1}_{j_1}}^{(0)}(0,0)\of{\psi^{\w_4}V^{\w_4}_{j_4}}(1,1) \of{\psi^{r,\w}V^{\w}_{j,m-1-r}}(\infty,\infty)\rangle  \\
    & = n_5^2C(j_1,j_4,j)Q_l^{2(m-j_1-j_4+1)}|\w_4^{j-m}b_1^{r+1}|^2 \nn \\
    & \times \int d^2y \Bigg|y^{j-m+r}(1-b_4 y)^{-j-j_1+j_4}(1+b_1 y)^{-j-j_4+j_1}\off{(j+j_4-j_1)\of{\frac{1-b_4 y}{1+b_1y}}+A_r}\Bigg|^2\,,
    \nn
\end{align}
and
\begin{align}
    &\langle \of{\psi^{r,\w}V^{\w}_{1-j,m-1-r}}\of{\psi^{\w_2}V^{\w_2}_{j_2}} \of{\psi^{\w_3} V^{\w_3}_{j_3}}^{(0)}\rangle = n_5^2C(1-j,j_2,j_3)Q_r^{2(j_2+m-j_3-1)}\\
    &\hspace{2.5cm} \times\int d^2u \Bigg|u^{1-j-m+r}(1-u)^{j-j_2+j_3-1}\off{(1-j+j_2-j_3)\frac{1}{1-u}+B_r}\Bigg|^2 \, , \nn
\end{align}
where the coefficients $A_r$ and $B_r$ are defined as 
\begin{align}
\begin{aligned}
    &A_r=\of{2j_1+n_5\w-2,2(j_1-1)+\frac{1}{b_1}(1-j_1+\frac{n_5}{2}\w_1),2(j_1-1)\frac{b_4}{b_1}}  \,,\\
    &B_r=\of{2 j_3 + n_5 w_3 - 2, j_3 + \frac{n_5}{2} w_3 - 1,0}\,,\quad 
\end{aligned}
\end{align}
with $r=-,0,+$.
Note that here we have introduced the notation 
\begin{equation}
   \of{\psi^{\w_i} V^{\w_i}_{j_i}}^{(0)}=\off{j^{-}_{-1-\w_i}-(h_i-1)\hat\jmath_{-1-\w_i}}\hat\psi^{\w_i} V^{\w_i}_{j_i}\,.
\end{equation}
for the SL(2,$\R$) part of the picture zero operators. 
It follows that, up the overall factors, 
\begin{align}
        &\langle\of{\psi^{\w_1} V^{\w_1}_{j_1}}^{(0)}\of{\psi^{\w_4}V^{\w_4}_{j_4}} \of{\psi^{r,\w}V^{\w}_{j,m-1-r}}\rangle\langle \of{\psi^{r,\w}V^{\w}_{1-j,m-1-r}}\of{\psi^{\w_2}V^{\w_2}_{j_2}} \of{\psi^{\w_3} V^{\w_3}_{j_3}}^{(0)}\rangle \sim \nn \\
    &\int d^2u d^2y \Bigg|   
    u^{-j-m}(1-u)^{j-j_2+j_3-1} y^{j-m-1}(1+b_4 y)^{-j+j_4-j_1}(1-b_1 y)^{-j-j_4+j_1}  \nn \\
    &\hspace{1.7cm}\Bigg[ 
      A_rB_r+ B_r (j+j_4-j_1)
       \frac{1+b_4y}{1-b_1y}+ A_r(1-j+j_2-j_3)\frac{1}{1-u} \label{4 double integrals}\\
    &\hspace{1.7cm} + (j+j_4-j_1)(1-j+j_2-j_3)
    \frac{1}{1-u}\frac{1+b_4y}{1-b_1y}
    \Bigg]\Bigg|^2, \nn
\end{align}
In order to show that the relevant linear combination of such products matches the four-point function expansion, we will need to make  use of the identity  
\begin{align}
    &\int_{\CC}d^2X \Big|X^{j-m-1}(1-b_4 X)^{m+\beta-1}\pFq{2}{1}{j+\alpha,j+\beta}{2j}{X}\Big|^2 = \label{HypMainIdentity}\\
    & \quad \frac{\pi\gamma(2j)}{\gamma(j+\beta)\gamma(j-\beta)} \int d^2u d^2y \Big|u^{-j-m}(1-u)^{j-\beta-1}y^{j-m-1}(1-b_1 y)^{-j-\alpha}(1+b_4y)^{-j+\alpha}\Big|^2  \,, \nn
\end{align}
with $b_1+b_4=1$, see Appendix \ref{sec: app A}. We stress that, as in the rest of the paper, here we merely use the absolute value squared as a placeholder for multiplying the holomorphic and antiholomorphic sectors, which might come with different powers. In other words, the choices of $(\alpha,\beta)$ and $(\bar\alpha,\bar\beta)$ in Eq.~\eqref{HypMainIdentity} are arbitrary and independent from each other. Now, using this for the different terms appearing in Eq.~\eqref{4 double integrals}, together with the well-known identity 
\begin{equation}
    a\,\pFq{2}{1}{a+1,b}{c}{x} = x \,\pFq{2}{1}{a,b}{c}{x}'+a\,\pFq{2}{1}{a,b}{c}{x}\,,
\end{equation}
allows us write the three-point function products as linear combinations of hypergometric functions and their derivatives. The three possible polarizations are given by
\begin{align}
    &\langle\of{\psi^{\w_1} V^{\w_1}_{j_1}}^{(0)}\of{\psi^{\w_4}V^{\w_4}_{j_4}} \of{\psi^{\w}V^{\w}_{j,m-1-r}}\rangle\langle \of{\psi^{\w}V^{\w}_{1-j,m-1-r}}\of{\psi^{\w_2}V^{\w_2}_{j_2}} \of{\psi^{\w_3} V^{\w_3}_{j_3}}^{(0)}\rangle \nn\\[1ex]
    & = n_5^4\Cc(j) a^{2(m-j_1-j_4-1)}\label{psi minus long}\\
    & \times \int d^2X \Bigg|X^{j-m+r}(1-b_4X)^{m-j_1-j_4-r}\Big[P^r_1(X){}_2F_1[X]+P^r_2(X) {}_2F'_1[X]+P^r_3(X){}_2F_1''[X]\Big]\Bigg|^2\nn \,.
\end{align}
Here the polynomials $P_i^r(X)$ depend on the choice of polarization, $r=-,0,+$. They take the form
\begin{align}
    P_1^{r}(X) = \Big[&(j_1+j_4+j+n_5\w-2)(j+j_2+j_3+n_5\w_3-2-(j-j_3+j_2)b_4X),\nn\\
    &(1-j_1+b_1(-2+j+j_1+j_4+\frac{n_5}{2}\w))(1-j-j_2-\frac{n_5}{2}\w_3-(j-j_3+j_2)b_4X),\nn\\
    &(j-j_1+j_4-b_4(j+j_1+j_4-2))(j-j_3+j_2) \Big]\,,\nn\\
    P_2^r(X)=\Big[&(2j+2j_3+n_5(\w+\w_3)-1-(1+2j+n_5\w)b_4X)X,\\
    &(j_1-1-b_1(2j+j_3+\frac{n_5}{2}(\w+\w_3))-(j_1-1-b_1(2j+1+\frac{n_5}{2}\w))b_4X)X ,\nn\\
    &(3+2j-2j_1-b_4(1+2j))X\Big]\nn\\
    P_3^r(X)=\Big[&X^2(1-b_4 X),b_1 X^2(b_4 X-1),b_1 X^2\Big]\,. \nn 
\end{align}
This shows that the four-point function \eqref{LongCont} can be written as
\begin{align}
     \mathbb{F}^{\ww}_{\CC}(x)= \int_{\frac{1}{2}+i\RR}dj |x|^{2(H-H_1-H_4)}&\sum_{r,\bar r}|a_{r}|^2\langle\Vv^{\w_1,(0)}_{j_1}\Vv^{\w_4}_{j_4} \Vv^{\w}_{j,r,\bar{r}}\rangle \langle\Vv^{\w}_{1-j,r,\bar{r}} \Vv^{\w_2}_{j_2}\Vv^{\w_3,(0)}_{j_3}\rangle\,,
     \label{linear combination}
\end{align}
where we have reinserted the antiholomorphic index $\bar{r}$ momentarily. Notice that while the bosonic propagators is taken into account by the reflection $j \to 1-j$, the fermionic two-point functions are important since they fix the relative coefficients, $a_r = (1,-2,1)$. The spacetime conformal weight of the intermediate state is
\begin{equation}
    H = m-1+\frac{n_5}{2}\w = \frac{-j(j-1)}{n_5\w}+\frac{(j_1+j_4-2)(j_1+j_4-1)}{n_5\w}+j_1+j_4-2+\frac{n_5}{2}\w \, .
\end{equation}
Hence, the spacetime $x$-dependence exactly matches the consistent behavior from the holographic CFT point of view.
This proves that the contributions to the four-point function  \eqref{Reg4pt} coming from the integral over the full complex plane can be  interpreted as a sum over the intermediate channels consistent with the spectrum and the selection rules of the model.  

\subsection{Other exchanges of short string states}\label{Sec: More short string channels}

For each possible AdS$_3$ polarization, short string states are given by Eq.~\eqref{LongString Vertex}, taking the vertex operator $V_{jm}^{\w}$ to lie in a flowed discrete representation of SL$(2,\RR)$. Therefore, the spin projection is parametrized as $m = j+p$, $p\in \NN$. These operators can be obtained from the $y$-basis ones as
\begin{equation}
    V_{j,j+p}^{\w}(x,z) = \lim_{y\to0}\partial_y^{p} V_j^{\w}(x,y,z).
\end{equation}
From the  point of view of the four-point function, these short string channels are encoded in the contribution from the small-$X$ region, see Eq.~\eqref{Reg4pt}. As discussed above, the integrand can be written as in \eqref{linear combination}, which allows us to distinguish between the contributions from different polarizations of the intermediate state.  
As an example, let us focus on the $\psi^\w\bar{\psi}^\w$-polarized channel (in our notation, this corresponds to $r=\bar r=-$). Its contribution gives 
\begin{align}
    &\text{P}_-=
    \int_{\frac{1}{2}+i\RR} dj |x|^{2(m-j_1-j_4+1)}n_5^4\mathfrak{C}(j)\int_{|X|<|x|} d^2X \Big|X^{j-m-1}(1-b_4X)^{m-j_1-j_4+1}\Big|^2\label{P-}
    \\&
    \quad \times \Big|\Big[(j_1+j_4+j+n_5\w-2)(j+j_2+j_3+n_5\w_3-2-(j-j_3+j_2)b_4X) {}_2F_1[X]\nn+\\
    &\quad (2j+2j_3+n_5(\w+\w_3)-1-(1+2j+n_5\w)b_4X)X {}_2F'_1[X]+X^2(1-b_4X){}_2F_1''[X]\Big]\Big|^2\nn \,.
\end{align}
Assuming $x<1$, Eq.~\eqref{P-} can be formally expanded in powers of $X$ as follows
\begin{align}
    \text{P}_-&=
    \int_{\frac{1}{2}+i\RR} dj |x|^{2(m-j_1-j_4+1)}n_5^4\mathfrak{C}(j)\int_{|X|<|x|} d^2X \sum_{p,\bar{p}=0}^{\infty}\Big|X^{j+p-m-1}
    A_-(p)\Big|^2\nn\,.
\end{align}
Using Eq.~\eqref{X power integral}, we first carry out the integral over $X$, which, at each order in $p$ (and $\bar{p}$), produces a simple pole in $j$, located at $j=m-p$, and further imposes the level matching constrain $p = \bar{p}$. Hence, after integrating over $j$ by means of \eqref{j power integral}, this leads to  
\begin{align}
    \text{P}_-&=
     \frac{2\pi^2 n_5^2\mathfrak{C}(j)}{2j+n_5\w-1}\sum_{p=0}^{\infty}\Big|x^{j+p-j_1-j_4+1}
    A_-(p)\Big|^2\nn\,.
\end{align}
At $j = m-p$, the coefficients $A_-(p)$ take the form 
\begin{align}
    A_-(p)=&(j-j_3+j_2)_p(j+j_2+j_3-2+n_5\w_3+p)(-b_4)^p\nn\\
    &\times\sum_{k=0}^p(-b_4)^{-k}\binom{p}{k}\frac{(j-j_1+j_4)_{k}}{(2j)_{k}}(j+j_1+j_4+n_5 \w-2+k)\,.
\end{align}
Therefore, after summing over $k$, we obtain
\begin{align}
    &\text{P}_-=
     \frac{2\pi^2 n_5^2\mathfrak{C}(j)}{2j+n_5\w-1}\times\\
     &\sum_{p=0}^{\infty}\Bigg|x^{j+p-j_1-j_4+1}
    (j-j_3+j_2)_p(j+j_2+j_3-2+n_5\w_3+p)(-b_4)^p(2j_1+n_5 \w-2)\nn\\
    &\Bigg(\pFq{2}{1}{-p,j-j_1+j_4}{2j}{\frac{1}{b_4}}+\frac{(j-j_1+j_4)}{(2j_1+n_5 \w-2)}\pFq{2}{1}{-p,j-j_1+j_4+1}{2j}{\frac{1}{b_4}}\Bigg)\Bigg|^2\nn\,.
\end{align}
On the other hand, at each value of $p$, the \textit{left} and \textit{right} short string three-point functions are given by
\begin{align}
    &\langle\of{\psi^{\w_1} V^{\w_1}_{j_1}}^{(0)}(0,0)\of{\psi^{\w_4}V^{\w_4}_{j_4}}(1,1) \of{\psi^{\w}V^{\w}_{j,j+p}}(\infty,\infty)\rangle =\\
    &C(j_1,j_4,j)b_4^{2p}\frac{\gamma(-p)\gamma(2j+p)}{\gamma(2j)}\Big|(2j_1+n_5\w-2)\pFq{2}{1}{-p,j-j_1+j_4}{2j}{\frac{1}{b_4}}\nn\\
    & \hspace{6cm}+(j+j_4-j_1)\pFq{2}{1}{-p,j-j_1+j_4+1}{2j}{\frac{1}{b_4}}\Big|^2\nn\,,
\end{align}
and 
\begin{align}
\begin{aligned}
    &\langle \of{\psi^{\w}V^{\w}_{j,j+p}}\of{\psi^{\w_2}V^{\w_2}_{j_2}} \of{\psi^{\w_3} V^{\w_3}_{j_3}}^{(0)}\rangle = \\
    &\hspace{3cm}C(j,j_2,j_3)\gamma(-p)|(j-j_3+j_2)_p(j+j_2+j_3+n_5\w_3+p-2)|^2\,,
\end{aligned}
\end{align}
respectively. Here we have used the identity 
\begin{align}
\Bigg|\pFq{2}{1}{-p,j-j_1+j_4}{2j}{\frac{1}{b_4}}\Bigg|^2 &=\\&\hspace{-1cm}\frac{\gamma(2j)}{\gamma(-p)\gamma(2j+p)}\int d^2 y \Big|y^{-p-1}\of{1-y}^{j_4-j-j_1}
\of{1+\frac{b_1}{b_4}y}^{j_1-j-j_4}\Big|^2\,,\nn
\end{align}
which holds inside the $j$-integral and for positive integer values of $p$, see Appendix \ref{sec: app A}.
It follows that,  as expected, the coefficient $A_-(p)$ matches the product 
\begin{equation}
     A_-(p) \sim \frac{\langle\of{\psi^{\w_1} V^{\w_1}_{j_1}}^{(0)}\of{\psi^{\w_4}V^{\w_4}_{j_4}} \of{\psi^{\w}V^{\w}_{j,j+p}}\rangle\langle \of{\psi^{\w}V^{\w}_{j,j+p}}\of{\psi^{\w_2}V^{\w_2}_{j_2}} \of{\psi^{\w_3} V^{\w_3}_{j_3}}^{(0)}\rangle}{\langle \of{\psi^{\w}V^{\w}_{j,j+p}}\of{\psi^{\w}V^{\w}_{j,j+p}}\rangle}.
\end{equation}

The remaining channels, including all other (holomorphic and antiholomorphic) polarizations can be studied analogously. It follows that small $X$ contribution to the complete four-point function reads 
\begin{align}
     \mathbb{F}^{\ww}_{|x|}(x)&=\int_{\frac{1}{2}+i\RR}dj|x|^{2(m-j_1-j_4+1)}\mathfrak{C}(j,l)n_5^4\nn\\
     &\int_{|X|<|x|}d^2X \Big|\sum_{p=0}^{\infty}X^{j-m+p-1}A_{-}(p)-2X^{j-m+p}A_{0}(p)+X^{j-m+p+1}A_{+}(p)\Big|^{2},\label{Short string series}
\end{align}
with coefficients $A_r(p)A_{\bar{r}}(\bar p)$ matching the relevant product of three-point functions, polarized along $\psi^{r,\w}\bar{\psi}^{\bar{r},\w}$ and at (unflowed) descendant levels $p,\bar{p}$. Note that, even though in general $p$ and $\bar{p}$ may not coincide, the level matching condition imposed by the integration over $X$ always gives $H=\bar{H}$, i.e.~we only deal with operators corresponding to spacetime scalars. This establishes all short string factorization channels.

\section{An interesting family of non-extremal four-point functions}  
\label{sec: non-extremal}

As discussed in \cite{Cardona:2010qf}, a certain family of unflowed, non-extremal correlators satisfying \eqref{Sphere condition} matches the holographic predictions from the boundary theory at the  symmetric orbifold point. We now show that this unexpected agreement between seemingly unprotected correlators computed at different points of the moduli space extends to the spectrally flowed sector. The worldsheet correlators of interest are given by
\begin{equation}
    \langle \Ww^{\w,(0)}_{j_1}(0,0) \Vv_{j_2}(1,1)\Vv^{\w\dagger,(0)}_{j_3}(\infty,\infty)\Vv^\dagger_{j_4}(x,z)\rangle.
    \label{non-extremal corr 1}
\end{equation}
Even though such correlators are non-extremal, they do satisfy the constrain \eqref{ExtremalJcondition}, implying \eqref{Sphere condition} holds. This means that only genus zero coverings contribute non-trivially to the HCFT correlator, hence we can restrict to the sphere topology on the string side. 

Let us compute this worldsheet correlation function. Fermion number counting in each sector leads to
\begin{align}
    &\langle \Ww^{\w,(0)}_{j_1}(0,0) \Vv_{j_2}(1,1)\Vv^{\w,(0)}_{j_3}(\infty,\infty)\Vv_{j_4}(x,z)\rangle=\\
    &\langle \Bb^{\w,1}_{j_1}(0,0) \Vv_{j_2}(1,1)\Aa^{\w,1}_{j_3}(\infty,\infty)\Vv_{j_4}(x,z)\rangle+\langle\Bb^{\w,2}_{j_1}(0,0) \Vv_{j_2}(1,1)\Aa^{\w,2}_{j_3}(\infty,\infty)\Vv_{j_4}(x,z)\rangle\nn
\end{align}
where $\Aa^{\w,i}_j$ and $\Bb^{\w,i}_j$ were defined in Eqs.~\eqref{Vpicture0} and \eqref{Wpicture0}.
The additional SL$(2,\RR)$ fermion in the second term makes it subleading in the expansion in powers of $x$.  The relevant bosonic SU(2) correlator is of the form
\begin{align}
    &\langle\of{K_{-1-\w}W^\w_{j_1}}W_{l_2}W^{\w,\dagger}_{l_3}W^{\dagger}_{l_4}\rangle =\\
    &\hspace{3cm}- \sum_l \Cc'(l)|z^{\Delta'-\Delta'_1-\Delta'_4-\w l_4 +1}(l-l_1+l_4)|^2\delta(l+l_2-l_3)\delta(l-l_1+l_4-1).\nn
\end{align}
This is obtained combining the SU(2) series identifications with the results of Appendix \ref{sec: App B - The SU(2) sector}. It follows that the sum over $l$ localizes on the contribution of an intermediate state with SU(2) spin $l = l_1-l_4+1=l_3-l_2$. Moreover, due to the chirality structure of \eqref{non-extremal corr 1}, SU$(2)$ charge conservation condition gives
\begin{equation}
    j_1-j_4+1 = j_3-j_2.
\end{equation}
Combined with \eqref{ExtremalJcondition}, this implies that $j_4 = 3/2$. This reflects the fact that, in the D1D5 CFT it was found that  only one operator runs in the intermediate channel  \cite{Pakman:2009ab}.

We will focus on the short-string contributions for simplicity. The remaining contributions from the SL(2,$\R$) sector can be studied as in previous sections. The spacetime correlator is obtained by integrating over $z$. We define $u = x/z^\w$, and consider the integral over the $|u|<|x|$ region. We find that, at leading order,  
\begin{align}
    &\mathbb{F}^{\ww}_{|u|<|x|}(x)=\int_{\frac{1}{2}+i\RR} dj |x|^{2(\frac{\Delta+\Delta'}{\w}-2(j_4-1))}\\
    &\qquad \times \int_{|u|<|x|} d^2u \Cc(j)\Cc'(l)\Big|u^{-3+j-j_1+j_4-\frac{\Delta+\Delta'}{\w}}(j+\frac{n_5\w}{2}+h_2+h_3-2)(l-l_1+l_4)\Big|^2\nn \, .
\end{align}
Integrating over $u$ therefore adds an extra pole in $j$, located at 
\begin{align}
    \frac{\Delta+\Delta'}{\w} +j-j_1+j_4-2=0 \, .
\end{align}
This is precisely Virasoro condition for chiral primary states, solved by $j= j_1-j_4+2 = l+1$. Therefore, after introducing the appropriate normalization factors, the four-point function becomes
\begin{align}
    &\mathbb{F}^{\ww}_{|u|<|x|}(x)=\frac{1}{N}\frac{(2h_3-1)^{3/2}}{2(h_3-1)(2h_3-3)^{1/2}}|x|^{-2}.
\end{align}
Here we have used that $h_3=j_3+\frac{n_5}{2}\w$ and $j_4=j_2=3/2$, while $j_1=j_3-1$. This reproduces the large $N$ holographic result given in Eq.~\eqref{holographic non extremal} exactly. 

Given that this unexpected holographic matching involving non-protected four-point functions extends to all flowed sectors, it is natural to conjecture that this family of correlators should be protected by supersymmetry. It would be interesting to formally prove this by working within the SCFT defined on the boundary of AdS$_3$. 


\section{Concluding remarks and outlook}
\label{sec: conclusions}
Spectral flow notoriously complicates the computation of worldsheet correlators for strings in AdS$_3\times S^3 \times T^4$. In the tensionless limit, the worldsheet SCFT simplifies \cite{Dei:2020zui,Dei:2023ivl,Eberhardt:2025sbi}, allowing for the exact computation of all $n$-point correlation functions, including the integration over the insertion points. In short, correlators localize on specific configurations which allow for the existence of holomorphic branched covering maps from the worldsheet to the AdS$_3$ boundary. This leads to an exact matching with the correlation functions of the holographically dual theory, namely the symmetric orbifold CFT which has the $\mathcal{N}=(4,4)$ $T^4$ model as its seed.

Away from this limit, both the spacetime theory and the worldsheet CFT become considerably more complicated. While the former is believed to be a marginal deformation of a symmetric orbifold with a non-compact seed, correlation functions of the latter do not localize anymore. They do, however, present simple poles at analogous locations on the string moduli space, and the corresponding residues can, in principle, be computed using free-field methods \cite{Knighton:2023mhq,Knighton:2024qxd,Sriprachyakul:2024gyl}. In this more complicated setting, only two- and three-point functions have been obtained exactly for arbitrary spectral flow charges \cite{Dei:2021xgh, Bufalini:2022toj}.

Importantly, a highly nontrivial relation between spectrally flowed bosonic four-point functions and their unflowed counterparts was further conjectured in \cite{Dei:2021yom}. In this paper, we have explored the implications of this proposal in the supersymmetric context. To this end, we have combined the methods of \cite{Iguri:2022pbp,Iguri:2023khc} for computing generic supersymmetric three-point functions (and in particular for those involved in the AdS$_3$/CFT$_2$ chiral ring) with the conformal block expansion of bosonic four-point functions discussed recently in \cite{Iguri:2024yhb}. Focusing on correlators conserving the spectral flow charge, we have studied the $x$- and $z$-dependence of the proposed four-point function. Here, both worldsheet and spacetime dependence are encoded in the polynomials that constrain the existence of the branched covering maps. A detailed numerical study of these polynomials provides the starting point for all subsequent computations.

We first considered the bosonic AdS$_3$ model.
We have shown that the integration over the worldsheet modulus cleanly separates into contributions from short and long strings. We found that this leads to a consistent spacetime OPE, where the conformal weight of the exchanged state satisfies the constraints imposed by the worldsheet Virasoro condition.
For extremal correlators in the supersymmetric AdS$_3\times$S$^3\times$T$^4$ model, we have proved that the spectrally flowed short-string sector completely reproduces the single-cycle contributions to the spacetime chiral ring, as calculated in \cite{Pakman:2009ab}. This match holds for all allowed values of the spectral flow, relies on the precise pole cancellation between the SL(2,$\mathbb{R}$) and SU(2) sectors \cite{Dabholkar:2007ey}, and incorporates the correct holographic normalization \cite{Maldacena:2001km,Iguri:2023khc}. This completes the study of the chiral sector of the boundary theory from the worldsheet perspective, except for the missing operators discussed in \cite{Seiberg:1999xz}.

We also went beyond the protected sector. More explicitly, we found that the contributions from long string states to protected four-point functions, and also the short string contributions to a series of non-protected correlators, precisely reproduce the  expected structure of the boundary OPE. We obtained these additional results mainly by making use of the techniques developed in \cite{Iguri:2024yhb}. 

The unexpected holographic matching derived for the non-extremal cases, which extends the results of \cite{Cardona:2010qf}, suggests that a non-renormalization theorem should hold for this specific family of correlators. This can potentially be explored in the near future.

More generally, our study of non-protected processes should help to understand how the spacetime theory behaves away from the $n_5=1$ point, which is \textit{not} Sym$^N$($T^4$). As in the bosonic case, this is expected to correspond to a marginal deformation of a different symmetric orbifold, where the seed theory includes a linear dilaton factor. The latter is holographically related to the radial direction of AdS$_3$. However, the precise definition of the theory remains slightly unclear \cite{Sriprachyakul:2024gyl,Yu:2024kxr,Yu:2025qnw,Nairz:2025kou}. The study of more general non-extremal correlators in the supersymmetric worldsheet theory along the lines of the present work, and in particular for correlators where spectral flow is not conserved, would shed light on the corresponding marginal deformation. 

On the other hand, and coming back to the bosonic model, our results and those of \cite{Iguri:2024yhb} provide strong evidence for the validity of the conjecture of \cite{Dei:2021yom}, which nevertheless remains unproven. It would be extremely interesting to derive this formula, perhaps along the lines of \cite{Bufalini:2022toj}. 

Beyond the present model, computational techniques developed in the context of the worldsheet SL(2,$\RR$) correlators have proven useful in other related contexts. This includes the holographic description of  $T\bar{T}$ deformations \cite{Smirnov:2016lqw,Cavaglia:2016oda,Baggio:2018rpv,Chang:2018dge}, in particular in their so-called  \textit{single-trace} incarnation \cite{Kutasov:2001uf,Israel:2003ry,Giveon:2017nie,Asrat:2017tzd,Giveon:2017myj,Giribet:2017imm,Chakraborty:2019mdf,Apolo:2019zai,Martinec:2021vpk,Cui:2023jrb,Georgescu:2024iam,Dei:2024sct,Dei:2025ilx}, which represent a rare example of non-AdS holography under considerable computational control.
There are also interesting applications for black hole phenomenology \cite{Witten:1991yr,Banados:1992wn,Maldacena:2000kv} in the context of the Fuzzball program \cite{Lunin:2001jy,Bena:2022rna}, where one can view the super WZW models we have considered in this paper as building blocks for  coset models associated with a certain family of black hole microstates \cite{Martinec:2017ztd,Martinec:2018nco,Martinec:2019wzw,Martinec:2020gkv,Martinec:2022okx,Bufalini:2021ndn,Bufalini:2022wyp,Bufalini:2022wzu,Dei:2024uyx,Martinec:2025xoy}. The techniques developed here allow for a deeper study of string dynamics in these highly non-trivial geometries.

\acknowledgments

The work of the authors is supported by CONICET-Argentina.  

\appendix 
\section{Some useful integrals over the complex plane}
\label{sec: app A}

Here we compute several integrals used in the main text. Recall that we have defined  $\gamma(z) = \Gamma(z)/\Gamma(1-\bar{z})$. We are interested in situations where $z$ and $\bar{z}$ differ by an integer number $n$. The Euler inversion formula for the Gamma function then gives 
\begin{equation}
\label{Euler gamma}
\gamma(z)\gamma(1-\zb) = 1\,, \qquad 
\gamma(z)\gamma(1-z) = \gamma(\zb)\gamma(1-\zb) = (-1)^n \,. 
\end{equation}
This can be used to compute the complex generalization of the Beta function \cite{Dotsenko:1986ca,Maldacena:2001km}, namely 
\begin{equation}
\label{I0def}
    I_0(a,b) \equiv \int d^2u |u^{a-1} (1-u)^{b-1}|^2 =  \pi \, \frac{\gamma(a)\gamma(b)}{\gamma(a+b)} \, .
\end{equation}
Here, as in the main part of the paper, we slightly abuse the  notation: the integrand should not be interpreted as a squared absolute value, but rather as 
\begin{equation}
 u^{a-1} (1-u)^{b-1}\ub^{\ab-1}  (1-\ub)^{\bb-1}\,, \nn
\end{equation}
and $\ab$ and $\bb$ need not be the complex conjugates of $a$ and $b$, respectively, although they must satisfy $a-\ab \in \mathbb{Z}$ and $b-\bb \in \mathbb{Z}$. 

The integral \eqref{I0def} can be computed as follows. We first define  $u=u_1+i u_2$, and then rotate the contour of the $u_2$-integral close to the imaginary axis, so that $u_2\to ie^{-2i\vep}u_2$. Here $\vep$ is a positive number which will be taken to zero at the end. This leads to 
\begin{align}
\begin{aligned}
    I_0(a,b) =  \, i e^{-2i\vep} \int_{-\infty}^{+\infty} d^2u_1 \int_{-\infty}^{+\infty}  d^2u_2 & (u_1-u_2e^{-2i\vep})^{a-1}
    (1-u_1+u_2e^{-2i\vep})^{b-1}
    \\ 
    & \times (u_1+u_2 e^{-2i\vep})^{\ab-1}
(1-u_1-u_2e^{-2i\vep})^{\bb-1}.
\end{aligned}
\end{align}
In terms of $u_\pm=u_1\pm u_2$ and at first order in $\vep$, one gets 
\begin{align}
\begin{aligned}
    I_0(a,b) =  \frac{i}{2} \int_{-\infty}^{+\infty} d^2u_+   d^2u_- & [u_-+i\vep (u_+-u_-)]^{a-1}
    [1-u_--i\vep (u_+-u_-)]^{b-1} \\
    & \times [u_+-i\vep (u_+-u_-)]^{\ab-1}
    [1-u_++i\vep (u_+-u_-)]^{\bb-1}.
\end{aligned}
\end{align}
When $u_+$ is integrated over $(-\infty,0)$ and $(1,+\infty)$, the integral over $u_-$ vanishes since one can close the contour at infinity without picking up any residues. On the other hand, for $u_+\in (0,1)$ the contour of integration for $u_-$ runs above $0$ but below $1$. The latter integral can equivalently be taken to go from $+\infty$ to $1$ and back, turning around $u_-=1$ clockwise. Consequently, one finds 
\begin{align}
    \begin{aligned}
    I_0(a,b) & = -\sin[\pi b]  \int_1^{\infty} du_-u_-^{a-1}(u_--1)^{b-1} 
    \int_0^1 du_+ u_+^{\ab-1}(1-u_+)^{\bb-1} 
    \, .
    \end{aligned}
\end{align}
The remaining real integrals are of the beta-function type, and can easily be carried out. Making use of the identities in Eq.~\eqref{Euler gamma} then leads to \eqref{I0def}. 

We now move to the more general complex integral
\begin{align}
\begin{aligned}
    I_1(a,b,c) & \equiv   \int d^2u |u^{a} (1-u)^{b}(u-z)^{c}|^2 \, .
\end{aligned}
\end{align}
with  $0<z<1$. Proceeding as above one now obtains two non-trivial contributions, namely
\begin{align}
\begin{aligned}
    s(\cb)I_1(a,b,c) = -s(c)s(a)
    I_{-\infty}^0(a,b,c)I_{0}^z(\ab,\bb,\cb)  + s(\cb)s(b) I_{1}^{\infty}(a,b,c)I_{z}^1(\ab,\bb,\cb)\,, 
\end{aligned}
\end{align}
where we have introduced the shorthands
\begin{equation}
    s(a) \equiv \sin(\pi a) \qqquad 
    I_{\alpha}^\beta(a,b,c) = \int_{\alpha}^\beta dt |t|^a |t-1|^{b}|t-z|^c \, ,
\end{equation}
and similarly for the anti-holomorphic counterparts. 
We now make use of the hypergeometric integrals 
\begin{eqnarray}
    I_{1}^\infty(a,b,c) & = & \frac{\Gamma(-a-b-c-1)\Gamma(b+1)}{\Gamma(-a-c)}  {}_2F_1(-c,-a-b-c-1,-a-c,z) \,, \\
    I_{0}^z(a,b,c) & = & z^{1+a+c}\,\frac{\Gamma(a+1)\Gamma(c+1)}{\Gamma(a+c+2)}  {}_2F_1(-b,a+1,a+c+2,z) \,,
\end{eqnarray}
and of the identities 
\begin{eqnarray}
    s(a+c)I_{-\infty}^0(a,b,c) & = & - s(c) I_0^{z}(a,b,c)+s(b) I_1^{\infty}(a,b,c)  \,, 
    \\[1ex]
    s(a+c)I_{z}^1(a,b,c) & = & -s(a)I_{0}^z (a,b,c)- 
    s(a+b+c)I_{1}^\infty (a,b,c) \, . 
\end{eqnarray}
This leads to 
\begin{align}
\begin{aligned}
    I_1(a,b,c) &= 
    \frac{s(\ab)s(c)}{s(\ab+\cb)} I_{0}^z(a,b,c) I_{0}^z(\ab,\bb,\cb)   - 
    \frac{s(\ab)s(b)}{s(\ab+\cb)} I_{1}^\infty(a,b,c) I_{0}^z(\ab,\bb,\cb)
    \\
    & \quad - \frac{s(b)s(\ab+\bb+\cb)}{s(\ab+\cb)} I_{1}^\infty(a,b,c) I_{1}^\infty(\ab,\bb,\cb)  
    - \frac{s(b)s(\ab)}{s(\ab+\cb)} I_{1}^\infty(a,b,c) I_{0}^z(\ab,\bb,\cb)  \nn \, .
\end{aligned}
\end{align}
The crossed terms cancel. For the remaining ones, the holomorphic and anti-holomorphic contributions nicely combine into products of hypergeometric and $\gamma$-functions.  As a result, we obtain the following complex generalization of the integral formula for the hypergeometric function, 
\begin{align}
        &\frac{\gamma(c)}{\pi\gamma(b)\gamma(c-b)} \int d^2u | u^{b-1}(1-u)^{c-b-1}(1-zu)^{-a}|^2 
        \label{C3 generalizada}\\
        &= |{}_2F_1(a,b,c,z)|^2  
        + \frac{\gamma(c)\gamma(1+a-c)\gamma(1+b-c)}{\gamma(2-c)\gamma(a)\gamma(b)} |z^{1-c} {}_2F_1(1+a-c,1+b-c,2-c,z) |^2 \, .\nn
\end{align}
Once again, here the squared absolute values should not be taken literally; they simply denote the product of a holomorphic expression times its anti-holomorphic counterpart, where $a$, $b$ and $c$ are replaced by $\ab$, $\bb$ and $\cb$, respectively, with $a-\ab,\, b-\bb,\, c-\cb \in \mathbb{Z}$. In other words, \eqref{C3 generalizada} constitutes a generalization of Eq.~(C.3) in \cite{Maldacena:2001km}, which is valid for $(\ab,\bb,\cb) = (a,b,c)$. Note that Eq.~\eqref{C3 generalizada} reduces to \eqref{I0def} for $a=\ab=0$.

Finally, for $a= j+\alpha$, $b = j+\beta$ and $c = 2j$, the two terms on the RHS of Eq.~\eqref{C3 generalizada} are related by a reflection $j\rightarrow 1-j$, hence for any function $g(j)$ invariant under this reflection we can write 
\begin{align}
\begin{aligned}\label{A14}
   & \int_{\frac{1}{2}+i \RR_{+}} dj g(j) \frac{\pi \gamma(2j)}{\gamma(j+\beta)\gamma(j-\beta)}\int d^2t |t^{j+\beta-1}(1-t)^{j-\beta-1}(1-xt)^{-j-\alpha}|^2 \\ 
   & \qquad = \int_{\frac{1}{2}+i \RR} dj g(j)\Bigg|\pFq{2}{1}{j+\alpha,j+\beta}{2j}{x}\Bigg|^{2}.
\end{aligned}
\end{align}
A related identity is the Appell function-type integral, given by 
\begin{align}
    \int d^2u |u^{\alpha-1}&(1-u)^{\beta+\beta'-\alpha-1}(1-qu)^{-\beta}(1-pu )^{-\beta'}|^2=\nn\\
    &|(1-p)|^{-2\alpha}\int d^2t \Bigg|t^{\alpha-1}(1-t)^{\beta+\beta'-\alpha-1}\left(1-\frac{q-p}{1-p}\,t\right)^{-\beta}\Bigg|^2\label{Appell},
\end{align}
which follows by setting 
\begin{equation}
    u = \frac{t}{1+p(t-1)}
\end{equation}
in \eqref{A14}. Assuming $g(j) = g(1-j)$ as before, this implies 
\begin{align}
    \int_{\frac{1}{2}+i \RR} dj g(j)\Bigg|\pFq{2}{1}{j+\alpha,j+\beta}{2j}{\frac{q-p}{1-p}}\Bigg|^{2} = \label{HypIden}\\ \int_{\frac{1}{2}+i \RR_+} dj \frac{\pi \gamma(2j)|(1-p)|^{2(j+\beta)}}{\gamma(j+\beta)\gamma(j-\beta)}\int d^2u |u^{j+\beta-1}&(1-u)^{j-\beta-1}(1-qu)^{-(j+\alpha)}(1-pu )^{-(j-\alpha)}|^2\nn \, .
\end{align}

 \section{Contributions from the supersymmetric SU(2) sector}\label{sec: App B - The SU(2) sector}

In the SU(2) WZW model, spectral flow merely reshuffles the spectrum, hence flowed correlators are linear combinations of the unflowed ones, including those with descendant insertions. However, we can conveniently derive all correlators relevant for our computation using techniques completely analogous to those developed for SL(2,$\R$) \cite{Eberhardt:2019ywk,Dei:2021xgh,Bufalini:2022toj}. Indeed, correlation functions of flowed SU(2) primaries also satisfy local Ward identities that turn into recursion relations analogous to those of the SL(2,R) model, provided one replaces $k \to - k '$ and $j_i \to - l_i$  \cite{Iguri:2023khc}. Similar statements can be made about the SL(2,$\R$) and SU(2) fermionic four-point functions. 

In this work, we make use of the spectrally flowed SU(2) four-point function in the coincidence limit, namely 
\begin{align}
\begin{aligned}
&\langle W_{l_1,-l_1}^{w_1}(0)W_{l_2,-l_2}^{w_2}(1)W_{l_3,-l_3}^{\dagger w_3}(\infty)W_{l_4,-l_4}^{w_4}(z) \rangle=\\
&\quad\quad 
\big|z^{\frac{k' w_1 w_4}{2}+w_1l_4+w_4l_1}(1-z)^{\frac{k' w_2 w_4}{2}+w_2l_4+w_4l_2}\big|^2 \int \prod_{i=1}^4 d^2 y_i | y_i^{-1}\\
&\quad \times y_{12}^{l_1+l_2-l_3+l_4}(1-y_1y_3)^{l_1-l_2+l_3-l_4}(1-y_2y_3)^{-l_1+l_2+l_3-l_4}(1-y_4y_3)^{2l_4}|^2\\[1ex]
&\quad\times  \left\langle W_{l_1}^0(0,0) W_{l_2}^0(1,1) W_{l_3}^0(\infty,\infty) W_{l_4}^0\left(\frac{(1-y_2 y_3) \, y_{41}}{y_{21}\, (1-y_4y_3)},z\right) \right \rangle \ .
\label{SU2 coincidence bosonic}
\end{aligned}
\end{align}
Rescaling each $y_i \to y_i/y_3$ for $i\neq 3$ shows that the integral over $y_3$ imposes the charge conservation condition
\begin{equation}
        l_3=l_1+l_2+l_4. \label{SU2 charge conservation}
\end{equation}
Moreover, the unflowed four-point function can be expanded as 
\begin{align}
    \langle& W_{l_1}^0(0,0) W_{l_2}^0(1,1) W_{l_3}^0(\infty,\infty) W_{l_4}^0\left(v,z\right) \rangle = \\
    & \quad \sum_{l=\max(|l_3-l_2|,|l_1-l_4|)}^{\min(l_3+l_2,l_1+l_4)}\Cc'(l_i,l)v^{l_1+l_4-l}z^{\Delta'(l)-\Delta'_1-\Delta'_4}\pFq{2}{1}{l_1-l_4-l,l_3-l_2-l}{-2l}{v}.\nn
\end{align}
 where we are using a short-hand notation for the product of unflowed three-point functions, 
\begin{equation}
  \Cc'(l_i,l)= C'(l_1,l_4,l)C'(l,l_2,l_3).
\end{equation}
Only the term with $l =l_1+l_4=l_3-l_2$ contributes non-trivially, hence we obtain 
\begin{align}
&\langle W_{l_1,-l_1}^{w_1}(0)W_{l_2,-l_2}^{w_2}(1)W_{l_3,-l_3}^{\dagger w_3}(\infty)W_{l_4,-l_4}^{w_4}(z) \rangle=\\
&\hspace{3cm}
\big|z^{\Delta'(l)-\Delta'_1-\Delta'_4+\frac{k' w_1 w_4}{2}+w_1l_4+w_4l_1}(1-z)^{\frac{k' w_2 w_4}{2}+w_2l_4+w_4l_2}\big|^2\Cc'(l_i,l)\, .\nn
\end{align}
The fermionic correlators needed for the computation of \eqref{SUSY correlator} can also be obtained from the SL(2,$\R$) results by taking $k\to-2$, $j_2,j_4\to-1$ and $j_1,j_3\to 0$. It follows that  the full contribution from the supersymmetric SU(2) correlator \eqref{SU2 and ghost} takes the form 
\begin{equation}
\langle W_{l_1}^{w_1}(0)W_{l_2}^{w_2}(1)W_{l_3}^{\dagger w_3}(\infty)W_{l_4}^{w_4}(z) \rangle\langle\hat{\chi}^{\w_1}(0)\hat{\chi}^{\w_2}(1) \hat{\chi}^{\w_3}(\infty)\hat{\chi}^{\w_4}(z)\rangle=z^{\boldsymbol{\Delta}_l'^{\w}-\boldsymbol{\Delta}'^{\w_1}_1-\boldsymbol{\Delta}'^{\w_4}_4}\Cc'(l_i,l).\end{equation}
As usual, the overall $z$-power involves the worldsheet conformal dimensions $\boldsymbol{\Delta}'^{\w_i}_i = \Delta'_i-\w l_i +(k'+2) w^2/4$, and takes the expected value for an  intermediate state with spectral flow charge $\w=\w_1+\w_4$ and unflowed SU(2) spin $l=l_3-l_2$.

\bibliographystyle{JHEP}
\bibliography{refs}

\end{document}